\documentclass[final,3p,times,fleqn]{elsarticle}
\usepackage{mathtools,booktabs}
\usepackage{bm}
\usepackage{autobreak}
\usepackage{amsmath}
\usepackage{amsthm}
\usepackage{siunitx}
\usepackage{gensymb}
\usepackage{subfigure}
\usepackage{epstopdf}
\usepackage{color}
\usepackage{hyperref}

\biboptions{sort&compress}
\begin{document}
\begin{frontmatter}
	\title{A comparative study of two Allen-Cahn models for immiscible $N$-phase flows by using a consistent and conservative lattice Boltzmann method}
	\author[a,b,c]{Chengjie Zhan}
	%\ead{zhancj@hust.edu.cn}
	\author[a,b,c]{Xi Liu}
	%\ead{aubrey-xixi@126.com}
	\author[a,b,c]{Zhenhua Chai \corref{cor1}}
	\ead{hustczh@hust.edu.cn}
	\author[a,b,c]{Baochang Shi}
	%\ead{shibc@hust.edu.cn}
	\address[a]{School of Mathematics and Statistics, Huazhong University of Science and Technology, Wuhan 430074, China}
	\address[b]{Institute of Interdisciplinary Research for Mathematics and Applied Science, Huazhong University of Science and Technology, Wuhan 430074, China}
	\address[c]{Hubei Key Laboratory of Engineering Modeling and Scientific Computing, Huazhong University of Science and Technology, Wuhan 430074, China}
	\cortext[cor1]{Corresponding author.}
	\begin{abstract} 
		In this work, we conduct a detailed comparison between two second-order conservative Allen-Cahn (AC) models [\emph{Model A}: Zheng \emph{et al.}, Phys. Rev. E 101, 0433202 (2020) and \emph{Model B}: Mirjalili and Mani, (2023)] for the immiscible $N$-phase flows. Mathematically, these two AC equations can be proved to be equivalent under some approximate conditions. However, the effects of these approximations are unclear from the theoretical point of view, and would be considered numerically. To this end, we propose a consistent and conservative lattice Boltzmann method for the AC models for $N$-phase flows, and present some numerical comparisons of accuracy and stability between these two AC models. The results show that both two AC models have good performances in accuracy, but the \emph{Model B} is more stable for the realistic complex $N$-phase flows, although there is an adjustable parameter in the \emph{Model A}.
	\end{abstract}
	\begin{keyword}
		Model comparisons \sep Allen-Cahn models \sep $N$-phase flows \sep lattice Boltzmann method 
	\end{keyword}		
\end{frontmatter}
\section{Introduction}
The phase-field method with an order parameter introduced to distinguish different phases, has been widely used in the study of multiphase flows \cite{Anderson1998ARFM,Badalassi2003JCP,Kim2005JCP,Boyer2006MMAN} for its advantages in needless of explicit interface-tracking and volume conservation. In this method, the interface of different phases is considered to have a small but finite thickness, and the physical variables are smoothly changed across the interface. In the phase-field method, there are two popular models, i.e., the fourth-order Cahn-Hilliard (CH) equation \cite{Cahn1996EJAM} and the second-order conservative Allen-Cahn (AC) equation \cite{Chiu2011JCP,Sun2007JCP}, that are usually adopted. However, these models cannot be directly used to study some more complex problems in nature and engineering processes, where more than two phases are included, for instance, the enhanced oil recovery \cite{Alvarado2010Energies,Maghzi2012ETFS}, emulsion formation \cite{Utada2005Science,Titibocchi2021NC}, and additive manufacturing \cite{Korneev2020CAD}. Therefore, to extend the scope of application of the phase-field model to depict more realistic multiphase systems, the phase-field models for two-phase problems should be extended in some way for the $N$-phase flows ($N\geq3$), and simultaneously, the extended models should follow the consistency of reduction, i.e., they can reduce to the models for $N-M$ phase flows when $M$ phase ($M<N$) are absent in the system with $N$ phases. 

The CH equation is popular in the simulation of multiphase flows due to its properties in the volume conservation and thermodynamic consistency. Boyer \emph{et al.} \cite{Boyer2006MMAN,Boyer2010TPM} elaborately designed a consistent bulk free energy for the ternary CH model, which can reduce to the double-well structure for two-phase cases when one phase disappears. Combing the distinct features of the available models \cite{Boyer2006MMAN,Kim2005JCP}, Kim \cite{Kim2007CMAME} proposed a ternary CH model with the continuous surface tension formulation to account for the surface tension effect, which allows the extension to more than three components. Recently, Dong \cite{Dong2018JCP} presented a reduction-consistent and thermodynamically-consistent formulation for an isothermal mixture consisting of $N$ immiscible incompressible fluids, where the free energy density function is equivalent to the form originally suggested in Ref. \cite{Boyer2014MMMAS}. However, it is usually difficult or even impossible to maintain the boundedness of the order parameter in the fourth-order CH equation, and the shrinkage/expansion occurs and the volume of a specific phase enclosed by the interface would diffuse into another phase to restore the equilibrium hyperbolic tangent profile \cite{Yue2007JCP}. In this case, although the total mass of CH system is conserved, there is a mass leakage among different phases, which would effect the accuracy of the interface position \cite{Yue2007JCP,Li2016CNSNS}. 
 
On the other hand, the second-order conservative AC model is much simpler than CH equation, and has been another strategy to capture the phase interfaces in $N$-phase flows. Abadi \emph{et al.} \cite{Abadi2018JCP} extended the conservative AC equation to the ternary fluids system, and the corresponding model is symmetric with respect to the phases and reduction-consistent. However, the consistency of reduction can not be guaranteed for the multiphase system with more than three phases. Motivated by the work of Ref. \cite{Lee2015PASMA}, Zheng \emph{et al.} \cite{Zheng2020PREa} developed a reduction-consistent and conservative AC equation, where a modified Lagrange multiplier is adopted to satisfy the consistency of reduction for $N$-phase flows. More recently, Mirjalili and Mani \cite{Mirjalili2023} proposed an $N$-phase extension of second-order conservative AC equation \cite{Chiu2011JCP}, which is in a simple conservative form and satisfies the volume conservation, the consistency of reduction, the symmetry with respect to the phases, and the consistency of mass-momentum transport. 

Actually, it is difficult to obtain the analytical solutions of the CH and AC equations, and thus many numerical methods have been developed to solve the phase-field equations in the past years, such as finite-difference method \cite{Jacqmin1999JCP,Furihata2001NM,Lee2012CPC,Li2016JCP,Lee2016MCS}, finite-element method \cite{Zhang2010JCP,Hua2011JCP}, spectral method \cite{Liu2003PDNP,Yue2004JFM,He2007ANM}, and to name but a few. As a popular kinetic-theory based numerical approach, the lattice Boltzmann (LB) method has also become an efficiently numerical tool in the simulation of complex fluid systems \cite{Higuera1989EPL,Chen1998ARFM,Aidun2010ARFM}, especially the multiphase flows \cite{Gunstensen1991PRA,Shan1993PRE,Swift1995PRL}. Up to now, some phase-field-based LB models for $N$-phase flows have been developed, which can be divided into two types, i.e, the first one \cite{Liang2016PRE,Abadi2018PRE,Zheng2019PRE,Yuan2022PF} based on the CH equation, and the other \cite{Abadi2018JCP,Zheng2020PREb,Hu2020IJMF} based on the conservative AC equation. However, the fourth-order CH equation cannot be exactly recovered by the second-order LB models through Chapman-Enskog expansion, and most of these models have some problems in the study of multiphase flows with large density ratios, mass conservation and the consistency of reduction. 

From above review and discussion, one can find that there are two simple and efficient AC models with some good properties that can be used for the $N$-phase flows well, the first is that in Ref. \cite{Zheng2020PREa} and the second is the one in Ref. \cite{Mirjalili2023}. It should be noted that, however, the differences and relations as well as the advantages and/or disadvantages of these two models are unclear. The aim of this work is to give a comparison between them. To this end, we first compared these two AC equations in theory, and then developed a consistent and conservative LB method for the AC models for the further numerical comparisons. The remainder of this paper is organized as follows. In Sec. \ref{GoverEqs}, the consistent and conservative AC models for incompressible $N$-phase flows are introduced, and a theoretical comparison between them is also conducted. Then a consistent and conservative multiple-relaxation-time LB method is developed in Sec. \ref{LBMs}. In Sec. \ref{Numerical}, the numerical validations and comparisons are performed, and finally, some conclusions are summarized in Sec. \ref{Conclusion}. 

\section{The consistent and conservative Allen-Cahn models for immiscible $N$-phase flows}\label{GoverEqs}
\subsection{The conservative Allen-Cahn models for immiscible $N$-phase flows}
The general second-order conservative AC equation can be written into the following form,
\begin{equation}\label{eq-NACE}
	\frac{\partial\phi_p}{\partial t}+\nabla\cdot\left(\phi_p\mathbf{u}\right)=\nabla\cdot M\left(\nabla\phi_p-\mathbf{R}_p\right),\quad\text{for } 1\leq p\leq N,
\end{equation}
where $\phi_p$ is the volume fraction of phase $p$ in total volume, and satisfies the volume conservation condition $\sum_p\phi_p=1$. $\mathbf{u}$ is the fluid velocity, $M$ represents the mobility, $\mathbf{R}_p$ is the source term and has specific forms for different conservative AC equations.

For the AC model in Ref. \cite{Zheng2020PREa} (hereafter \emph{Model A}), $\mathbf{R}_p$ is given by
\begin{equation}\label{eq-source1}
	\mathbf{R}_p=\frac{4\phi_p\left(1-\phi_p\right)}{\epsilon}\mathbf{n}_p-\bm{\beta}_p,\quad\bm{\beta}_p=\frac{\phi_p^n}{\sum_{q}\phi_q^n}\sum_{q}\frac{4\phi_q\left(1-\phi_q\right)}{\epsilon}\mathbf{n}_q,
\end{equation}
where $\epsilon$ is the interface thickness, $\mathbf{n}_p=\nabla\phi_p/|\nabla\phi_p|$ is the normal vector of the $p$-th fluid interface, $\bm{\beta}_p$ is the Lagrange multiplier with $n$ being a positive parameter. 
This model is a direct extension of the AC equation for two-phase flow, and the first part of the source term $\mathbf{R}_p$ in Eq. (\ref{eq-source1}) is the same as the source term of two-phase AC equation \cite{Sun2007JCP,Chiu2011JCP}. The key point of \emph{Model A} is to introduce an appropriate $\bm{\beta}_p$ to guarantee the properties of the volume conservation and the consistency of reduction. 

For the AC model proposed in Ref. \cite{Mirjalili2023}, here an equal interface thickness parameter ($\epsilon$) between all phase pairs is considered (hereafter \emph{Model B}), $\mathbf{R}_p$ is designed as
\begin{equation}\label{eq-source2}
	\mathbf{R}_p=\sum_{q\neq p}\frac{4\phi_p\phi_q}{\epsilon}\mathbf{n}_{pq}.
\end{equation}
Here the pairwise normal vector $\mathbf{n}_{pq}$ is defined as
\begin{equation}
	\mathbf{n}_{pq}=\frac{\nabla\phi_{pq}}{|\nabla\phi_{pq}|},\quad\text{for } q\neq p,
\end{equation}
where $\phi_{pq}=\phi_p/\left(\phi_p+\phi_q\right)$ is the pairwise volume fraction. Obviously, $\phi_{pq}$ reduces to $\phi_p$ in the case of $\phi_p+\phi_q=1$. The idea of \emph{Model B} is that most regions of the $N$-phase system are actually two-phase system, therefore, the pairwise volume fraction $\phi_{pq}$ is introduced to design the source term $\mathbf{R}_p$ to satisfy the volume conservation, the symmetry with respect to the phases, and the consistency of reduction. Actually, one can easily find that both models are the same as each other and equal to the AC equation \cite{Chiu2011JCP} for two-phase flows when $N=2$.

To give a comparison of above two AC models when $N\geq3$, the equilibrium distributions of the phase-field variable are firstly introduced. The standard order parameter $\phi_p$ at equilibrium state usually satisfies a hyperbolic tangent profile: $\phi_p^{eq}=0.5+0.5\tanh\left(2x_p/\epsilon\right)$ with $x_p$ being the signed distance function. With this equilibrium distribution, the norm of the gradient of $\phi_p$ can be expressed by
\begin{equation}\label{eq-Dphipeq}
	|\nabla\phi_p|^{eq}=\frac{4\phi_p\left(1-\phi_p\right)}{\epsilon},\quad\text{for }1\leq p\leq N.
\end{equation}
Similarly, the equilibrium distribution of the pairwise volume fraction $\phi_{pq}$ is also expected to satisfy $\phi_{pq}^{eq}=0.5+0.5\tanh\left(2x_{pq}/\epsilon\right)$, and its gradient norm can be given by
\begin{equation}\label{eq-Dphipqeq}
	|\nabla\phi_{pq}|^{eq}=\frac{4\phi_{pq}\left(1-\phi_{pq}\right)}{\epsilon}=\frac{4\phi_p\phi_q}{\epsilon\left(\phi_p+\phi_q\right)^2},\quad\text{for }1\leq p\neq q\leq N.
\end{equation}

Subsequently, we can rewrite the source term $\mathbf{R}_p$ of the two AC models in terms of normal vectors $\mathbf{n}_p$ and $\mathbf{n}_q$ ($q\neq p$), and make a comparison to show their relation.

\noindent\emph{Model A}:
\begin{equation}\label{eq-RpA}
	\mathbf{R}_p^A=\frac{4\phi_p\left(1-\phi_p\right)}{\epsilon}\mathbf{n}_p-\frac{\phi_p^n}{\sum_{q}\phi_q^n}\sum_{q}\frac{4\phi_q\left(1-\phi_q\right)}{\epsilon}\mathbf{n}_q=\frac{\sum_{q\neq p}\phi_q^n}{\sum_{q}\phi_q^n}\frac{4\phi_p\left(1-\phi_p\right)}{\epsilon}\mathbf{n}_p-\frac{\phi_p^n}{\sum_{q}\phi_q^n}\sum_{q\neq p}\frac{4\phi_q\left(1-\phi_q\right)}{\epsilon}\mathbf{n}_q.
\end{equation}

\noindent\emph{Model B}:
\begin{equation}\label{eq-RpB}
	\mathbf{R}_p^B=\sum_{q\neq p}\frac{4\phi_p\phi_q}{\epsilon|\nabla\phi_{pq}|}\frac{\phi_q\nabla\phi_p-\phi_p\nabla\phi_q}{\left(\phi_p+\phi_q\right)^2}=\sum_{q\neq p}\frac{4\phi_p\phi_q^2|\nabla\phi_p|}{\left(\phi_p+\phi_q\right)^2|\nabla\phi_{pq}|}\mathbf{n}_p-\sum_{q\neq p}\frac{4\phi_p^2\phi_q|\nabla\phi_q|}{\left(\phi_p+\phi_q\right)^2|\nabla\phi_{pq}|}\mathbf{n}_q.
\end{equation}
In above equation (\ref{eq-RpB}), if the norms of gradient $|\nabla\phi_p|$ and $|\nabla\phi_{pq}|$ are approximated by the equilibrium distributions (\ref{eq-Dphipeq}) and (\ref{eq-Dphipqeq}), we have
\begin{equation}
	\mathbf{R}_p^B=\sum_{q\neq p}\frac{4\phi_q\phi_p\left(1-\phi_p\right)}{\epsilon}\mathbf{n}_p-\sum_{q\neq p}\frac{4\phi_p\phi_q\left(1-\phi_q\right)}{\epsilon}\mathbf{n}_q=\sum_{q\neq p}\phi_q\frac{4\phi_p\left(1-\phi_p\right)}{\epsilon}\mathbf{n}_p-\phi_p\sum_{q\neq p}\frac{4\phi_q\left(1-\phi_q\right)}{\epsilon}\mathbf{n}_q.
\end{equation}
In this case, if $n=1$ is considered in Eq. (\ref{eq-RpA}), one can find that these two source terms $\mathbf{R}_p^A$ and $\mathbf{R}_p^B$ are the same with each other, i.e., \emph{Model B} is equivalent to \emph{Model A} when $n=1$ and the approximations of the gradient norms. 

In summary, these two AC models are equivalent under some conditions. \emph{Model A} with an adjustable parameter $n$ looks more flexible than \emph{Model B}, while the optimal value of $n$ needs to be tested. On the other hand, the effect of the approximations of the gradient norms is not clear, which will be studied through the following numerical simulations.

\subsection{The consistent and conservative Navier-Stokes equations for incompressible flows}
To describe the fluid flows, the following consistent and conservative incompressible Navier-Stokes (NS) equations are used \cite{Huang2020JCPa,Mirjalili2021JCP,Zhan2022PRE},
\begin{subequations}\label{eq-NS}
	\begin{equation}
		\nabla\cdot\mathbf{u}=0,
	\end{equation}
	\begin{equation}\label{momentumEq}
		\frac{\partial\left(\rho\mathbf{u}\right)}{\partial t}+\nabla\cdot\left(\mathbf{m}\mathbf{u}\right)=-\nabla P+\nabla\cdot\mu\left[\nabla\mathbf{u}+\left(\nabla\mathbf{u}\right)^\top\right]+\mathbf{F}_s+\mathbf{F}_g,
	\end{equation}
\end{subequations}
where $\rho$ is the density, $\mathbf{m}$ is the mass flux, $P$ is the hydrodynamic pressure, $\mu$ represents the dynamic viscosity, $\mathbf{F}_s$ is the surface tension force, and $\mathbf{F}_g$ is the other external force such as gravity. 

The density $\rho$ in multiphase flows is dependent on the phase-field variable $\phi_p$, and is defined by
\begin{equation}\label{eq-rho}
	\rho=\sum_p\rho_p\phi_p,
\end{equation}
where $\rho_p$ is the density of component $p$. The viscosity is also assumed as a linear function of $\phi_p$ to smoothly across the interface,
\begin{equation}
	\mu=\sum_p\mu_p\phi_p,
\end{equation}
with $\mu_p$ being the dynamic viscosity of phase $p$.

To satisfy the consistency of mass conservation and the consistency of mass-momentum transport \cite{Huang2020JCPa,Mirjalili2021JCP}, the mass flux is defined as
\begin{equation}
	\mathbf{m}=\rho\mathbf{u}+\mathbf{m}_{\phi},\quad \mathbf{m}_{\phi}=-\sum_{p}\rho_pM\left(\nabla\phi_p-\mathbf{R}_{p}\right),
\end{equation} 
where the mass diffusion between different phases $\mathbf{m}_{\phi}$ is included. 

We now introduce the surface tension force. It is noted that to exactly compare these two AC models, the same potential form of the surface tension force is adopted, instead of the potential form in Ref. \cite{Zheng2020PREa} and the continuum surface force in Ref. \cite{Mirjalili2023}. 
In the phase-field theory, a reduction-consistent total free energy of multiphase fluids system can be written as \cite{Boyer2014MMMAS,Dong2018JCP}
\begin{equation}
	E=\int_{\Omega}\sum_{q\neq p}\left\{\beta_{pq}\left[g\left(\phi_p\right)+g\left(\phi_q\right)-g\left(\phi_p+\phi_q\right)\right]+\frac{k_{pq}}{2}\nabla\phi_p\cdot\nabla\phi_q\right\}d\Omega,\quad\text{for }1\leq p,q\leq N.
\end{equation}
Here $g\left(\phi\right)=\phi^2\left(1-\phi\right)^2$, $\beta_{pq}$ and $k_{pq}$ are two physical parameters related to the interface thickness $\epsilon$ and the pairwise symmetric surface tension coefficient $\sigma_{pq}$,
\begin{equation}
	\beta_{pq}=\frac{3}{\epsilon}\sigma_{pq},\quad k_{pq}=-\frac{3\epsilon}{4}\sigma_{pq},
\end{equation} 
where $\sigma_{pq}$ satisfies $\sigma_{pq}=\sigma_{qp}>0$ ($p\neq q$) and $\sigma_{pp}=0$.
From above expression of the free energy, the potential from of the surface tension force can be given by
\begin{equation}\label{eq-Fs1}
	\mathbf{F}_s=\sum_{p}\mu_{\phi_p}\nabla\phi_{p}, 
\end{equation}
where $\mu_{\phi_p}$ is the chemical potential of phase $p$,
\begin{equation}
	\mu_{\phi_p}=\frac{\delta E}{\delta\phi_p}=\sum_{q\neq p}2\beta_{pq}\left[g'\left(\phi_p\right)-g'\left(\phi_p+\phi_q\right)\right]-\sum_{q\neq p}k_{pq}\nabla^2\phi_q.
\end{equation}
	
\section{The consistent and conservative lattice Boltzmann method for immiscible $N$-phase flows}\label{LBMs}
In this section, we will develop a consistent and conservative LB method for the above AC models [Eqs. (\ref{eq-NACE}) and (\ref{eq-NS})].
We first propose the LB model for phase-field equation (\ref{eq-NACE}), and the evolution equation reads
\begin{equation}
	f_{i,p}\left(\mathbf{x}+\mathbf{c}_i\Delta t,t+\Delta t\right)=f_{i,p}\left(\mathbf{x},t\right)-\Lambda_{ij}^p\left[f_{i,p}\left(\mathbf{x},t\right)-f_{i,p}^{eq}\left(\mathbf{x},t\right)\right]+\Delta t\left(\delta_{ij}-\frac{\Lambda_{ij}^p}{2}\right)R_{j,p}\left(\mathbf{x},t\right),\quad\text{for }1\leq p\leq N,
\end{equation}
where $f_{i,p}\left(\mathbf{x},t\right)$ ($i=0,1,\cdots,Q-1$ with $Q$ being the number of the discrete velocity directions) represents the distribution function of phase-field variable $\phi_p$ at position $\mathbf{x}$ and time $t$, and $f_{i,p}^{eq}\left(\mathbf{x},t\right)$ is the corresponding equilibrium distribution function. $\mathbf{c}_i$ is the discrete velocity, $\Delta t$ is the time step, and $\bm{\Lambda}^p=(\Lambda_{ij}^p)$ represents the invertible collision matrix. The expression of the distribution functions $f_{i,p}^{eq}\left(\mathbf{x},t\right)$ and $R_{i,p}\left(\mathbf{x},t\right)$ can be designed as
\begin{equation}
	f_{i,p}^{eq}=\omega_{i,p}\phi_p\left(1+\frac{\mathbf{c}_i\cdot\mathbf{u}}{c_{s,p}^2}\right),\quad R_{i,p}=\omega_{i,p}\left[\frac{\mathbf{c}_i\cdot\partial_t\left(\phi_p\mathbf{u}\right)}{c_{s,p}^2}+\mathbf{c}_i\cdot\mathbf{R}_p\right],
\end{equation}
where $\omega_{i,p}$ is the weight coefficient and $c_{s,p}$ is the lattice sound speed. 

Similar to the LB models for AC equation of two-phase flows \cite{Wang2019Capillarity}, Eq. (\ref{eq-NACE}) can be correctly recovered through the direct Taylor (or Chapman-Enskog) expansion under the general LB framework \cite{Chai2020PRE} with the following relation,
\begin{equation}
	M=\left(\tau_f-0.5\right)c_{s,p}^2\Delta t,
\end{equation}
where $\tau_f$ is the relaxation parameter related to an eigenvalue of the collision matrix $\bm{\Lambda}^p$. Here we would also like to point out that the previous LB method in Ref. \cite{Zheng2020PREa} cannot give correct \emph{Model A} since there are some additional terms in the recovered macroscopic equation, which may also influence the numerical results \cite{Liang2023PRE}. 

The LB model for flow field is similar to that in our previous works \cite{Zhan2022PRE,Liu2022MMS}, here it is only briefly introduced. The evolution equation of the model can be expressed as
\begin{equation}
	g_i\left(\mathbf{x}+\mathbf{c}_i\Delta t,t+\Delta t\right)=g_i\left(\mathbf{x},t\right)-\Lambda_{ij}^g\left[g_i\left(\mathbf{x},t\right)-g_i^{eq}\left(\mathbf{x},t\right)\right]+\Delta t\left(\delta_{ij}-\frac{\Lambda_{ij}^g}{2}\right)F_j\left(\mathbf{x},t\right),
\end{equation}
where $g_i\left(\mathbf{x},t\right)$ is the distribution function for flow field, the distribution functions $g_i^{eq}\left(\mathbf{x},t\right)$ and $F_i\left(\mathbf{x},t\right)$ are given by
\begin{equation}
	g_i^{eq}=\lambda_i+\omega_i\left[\frac{\mathbf{c}_i\cdot\rho\mathbf{u}}{c_s^2}+\frac{\mathbf{m}\mathbf{u}:\left(\mathbf{c}_i\mathbf{c}_i-c_s^2\mathbf{I}\right)}{2c_s^4}\right],\quad \lambda_0=\left(\omega_0-1\right)\frac{P}{c_s^2}+\rho_0,\,  \lambda_i=\omega_i\frac{P}{c_s^2} (i\neq 0),
\end{equation}
\begin{equation}
	F_i=\omega_i\left[\mathbf{u}\cdot\nabla\rho+\frac{\mathbf{c}_i\cdot\mathbf{F}}{c_s^2}+\frac{\left(\mathbf{M}_{2F}-c_s^2(\mathbf{u}\cdot\nabla\rho)\mathbf{I}\right):\left(\mathbf{c}_i\mathbf{c}_i-c_s^2\mathbf{I}\right)}{2c_s^4}\right],
\end{equation}
where $\omega_i$ and $c_s$ are also the weight coefficient and lattice sound speed, which can be different from those of LB models for phase-field equations. The total external force $\mathbf{F}=\mathbf{F}_s+\mathbf{F}_g+\mathbf{F}_c$ with $\mathbf{F}_c=-\nabla\cdot\left(\mathbf{m}_{\phi}\mathbf{u}-\mathbf{u}\mathbf{m}_{\phi}\right)/2$ being a modified force term to correctly recover the consistent momentum equation (\ref{momentumEq}), the moment $\mathbf{M}_{2F}$ is designed as
\begin{equation}
	\mathbf{M}_{2F}=\partial_t\left(\rho\mathbf{u}\mathbf{u}+\frac{\mathbf{m}_{\phi}\mathbf{u}+\mathbf{u}\mathbf{m}_{\phi}}{2}\right)+c_s^2\left(\mathbf{u}\nabla\rho+\nabla\rho\mathbf{u}\right)+c_s^2\left(\mathbf{u}\cdot\nabla\rho\right)\mathbf{I}.
\end{equation} 

The phase-field variable, macroscopic velocity, and pressure are computed by
\begin{subequations}
	\begin{equation}
		\phi_p=\sum_if_{i,p},\quad\text{for }1\leq p\leq N,
	\end{equation}
	\begin{equation}\label{eq-u}
		\rho\mathbf{u}=\sum_i\mathbf{c}_ig_i+\frac{\Delta t}{2}\mathbf{F},
	\end{equation}
	\begin{equation}
		P=\frac{c_s^2}{1-\omega_0}\left[\sum_{i\neq 0}g_i+\frac{\Delta t}{2}\mathbf{u}\cdot\nabla\rho-\omega_0\frac{\mathbf{m}\cdot\mathbf{u}}{2c_s^2}+\omega_0\frac{\Delta t}{2}\frac{\partial_t\left(\mathbf{m}\cdot\mathbf{u}\right)}{2c_s^2}\right],
	\end{equation}
\end{subequations}
where the density $\rho$ is refreshed by Eq. (\ref{eq-rho}).
Finally, through the some asymptotic analysis methods, the NS equations (\ref{eq-NS}) can be recovered with the following relation, 
\begin{equation}
	\mu=\rho\left(\tau_g-0.5\right)c_s^2\Delta t,
\end{equation}
where $\tau_g$ is a relaxation parameter and will be stated below.  

Here it is noted that the velocity is implied in the right-hand-side of Eq. (\ref{eq-u}), and the value of the previous time step is adopted for simplicity. Additionally, the derivative terms in above LB method should be discretized with suitable difference schemes. In this work, $\partial_t\left(\rho\mathbf{uu}\right)$ is approximated by $\mathbf{u}\mathbf{F}+\mathbf{F}\mathbf{u}$, the explicit Euler scheme $\partial_t\chi\left(t\right)=\left[\chi\left(t\right)-\chi\left(t-\Delta t\right)\right]/\Delta t$ is adopted for the other temporal derivatives, and the following second-order isotropic central schemes are applied for the gradient and Laplacian operators,
\begin{equation}
	\begin{aligned}
		\nabla\chi\left(\mathbf{x}\right)&=\sum_{i\neq 0}\frac{\omega_i\mathbf{c}_i\chi\left(\mathbf{x}+\mathbf{c}_i\Delta t\right)}{c_s^2\Delta t},\\
		\nabla^2\chi\left(\mathbf{x}\right)&=\sum_{i\neq 0}\frac{2\omega_i\left[\chi\left(\mathbf{x}+\mathbf{c}_i\Delta t\right)-\chi\left(\mathbf{x}\right)\right]}{c_s^2\Delta t^2}.
	\end{aligned}
\end{equation}

\section{Numerical validations and discussion}\label{Numerical}	 
In this section, to validate the present LB method and to give a comparison between two AC models, several two-dimensional benchmark multiphase flow problems are considered here, including the static droplets, the spreading of a compound droplet on a solid wall, the spreading of a liquid lens, the Rayleigh-Taylor instability (RTI), and the dam break problems. Here the optimal value of $n$ in \emph{Model A} is not tested, and only two values of $n=1$ and 2 are adopted for simplicity. 
Moreover, it is worth noted that only $N-1$ phase-field equations are solved by the LB method, and the distribution of $N$-th phase can be obtained through the volume conservation condition.
In the following simulations, the D2Q9 lattice structure is adopted for all LB models, in which $c_s=c/\sqrt{3}$ with $c=\Delta x/\Delta t$. The half-way bounce-back scheme is applied for the no-flux scalar and no-slip velocity boundary conditions. To improve the efficiency of the LB algorithm, the collision step in LB method is carried out in the moment space through the transformation $\bm{\Lambda}=\mathbf{M}^{-1}\mathbf{SM}$ with $\mathbf{S}$ being a block-lower-triangle relaxation matrix \cite{Chai2020PRE}. In this work, the following natural transformation matrix $\mathbf{M}$ is considered:
\begin{equation}
	\mathbf{M}=\begin{pmatrix}
		1 &  1 &  1 &  1 &  1 &  1 &  1 &  1 &  1\\
		0 &  1 &  0 & -1 &  0 &  1 & -1 & -1 &  1\\
		0 &  0 &  1 &  0 & -1 &  1 &  1 & -1 & -1\\
		0 &  1 &  0 &  1 &  0 &  1 &  1 &  1 &  1\\
		0 &  0 &  1 &  0 &  1 &  1 &  1 &  1 &  1\\		
		0 &  0 &  0 &  0 &  0 &  1 & -1 &  1 & -1\\		
		0 &  0 &  0 &  0 &  0 &  1 &  1 & -1 & -1\\
		0 &  0 &  0 &  0 &  0 &  1 & -1 & -1 &  1\\
		0 &  0 &  0 &  0 &  0 &  1 &  1 &  1 &  1\\
	\end{pmatrix},
\end{equation}
and the relaxation matrix is set as a simple diagonal matrix $\mathbf{S}=\mathbf{diag}\left(s_0,s_1,s_1,s_2,s_2,s_2,s_3,s_3,s_4\right)$. Specifically, $s_1=1/\tau_f$ for phase field, while $s_2=1/\tau_g$ for flow field, and the others are set to be unity unless otherwise stated. 

In addition, to quantitatively evaluate the accuracy of the present LB method, the following $L^1$ norm of the relative error $E_{L^1}$ is applied:
\begin{equation}
	E_{L^1}=\frac{||\chi_a-\chi_n||_1}{||\chi_a||_1},
\end{equation}
where $\chi$ denotes a scalar characteristic quantity, and the subscripts $a$ and $n$ represent the analytical and numerical solutions of $\chi$. 

\subsection{Static droplets}\label{droplets}
\begin{figure}
	\centering
	\includegraphics[width=4.5in]{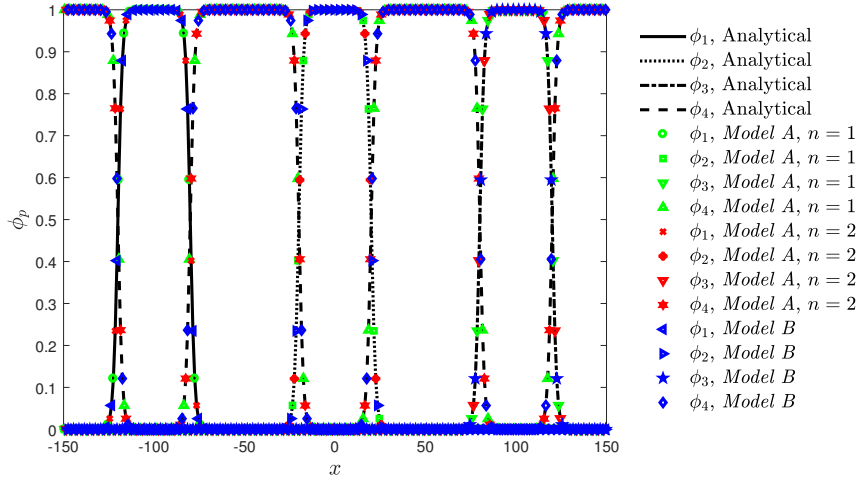}
	\caption{A comparison of the numerical and analytical profiles of the phase-field variables along the centerline of static droplets in the quaternary system.}
	\label{fig-droplets4}
\end{figure}
\begin{figure}
	\centering
	\subfigure[]{
		\begin{minipage}{0.48\linewidth}
			\centering
			\includegraphics[width=3.0in]{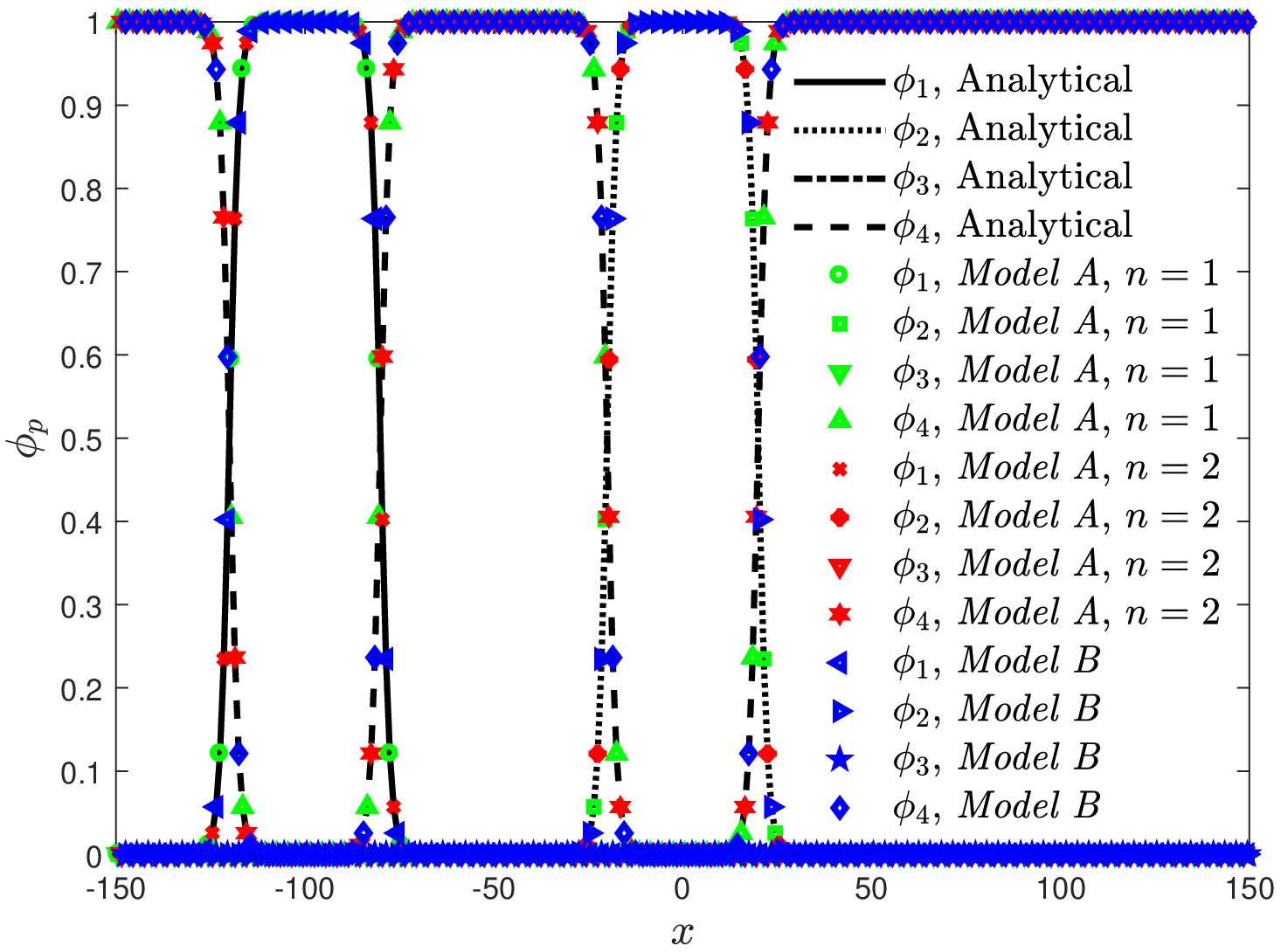}
	\end{minipage}}
	\subfigure[]{
		\begin{minipage}{0.48\linewidth}
			\centering
			\includegraphics[width=3.0in]{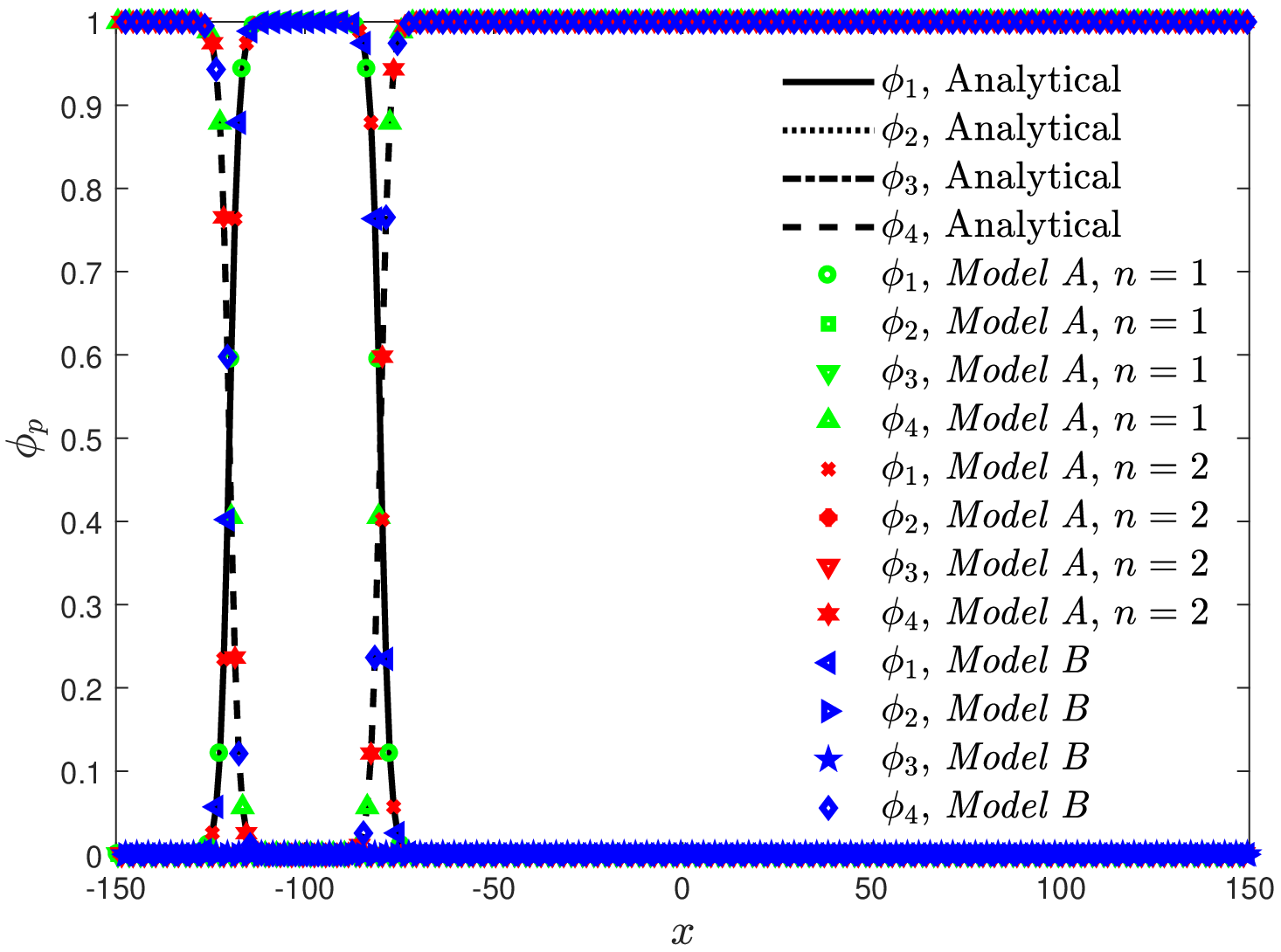}
	\end{minipage}}
	\caption{A comparison of the numerical and analytical profiles of the phase-field variables along the centerline of static droplets in (a) the ternary and (b) the binary systems.}
	\label{fig-droplets2}
\end{figure}
\begin{figure}
	\centering
	\includegraphics[width=3.5in]{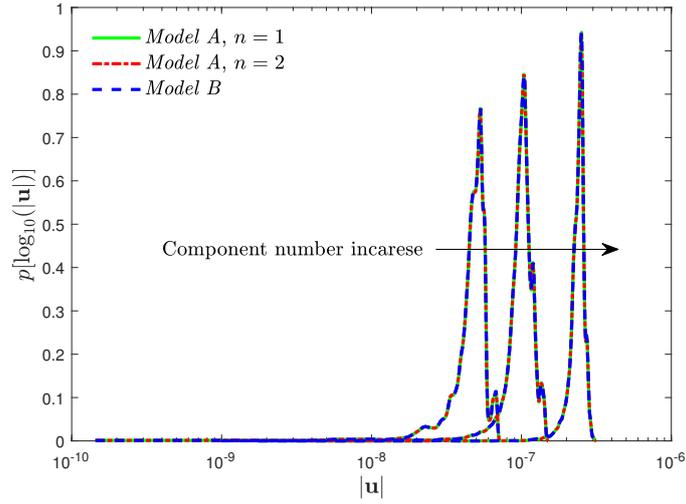}	
	\caption{The histograms of the logarithm of the velocity magnitude ($p[\log_{10}(|\mathbf{u}|)]$) of static droplets in different systems.}
	\label{fig-droplets-p}
\end{figure}

The simple problem of static droplets is firstly considered to test the present LB method. Initially, three same size droplets of different phases (phase 1, phase 2, and phase 3) with radius $R=20$, surrounded by phase 4, are symmetrically placed on the centerline of a domain $[-150,150]\times[-50,50]$. The periodic boundary condition is applied at all boundaries, and the initial distributions of the phase-field variables are given by
\begin{equation}
	\begin{aligned}
		\phi_1\left(x,y\right)&=0.5+0.5\tanh\frac{R-\sqrt{\left(x+100\right)^2+y^2}}{\epsilon/2},\\
		\phi_2\left(x,y\right)&=0.5+0.5\tanh\frac{R-\sqrt{x^2+y^2}}{\epsilon/2},\\
		\phi_3\left(x,y\right)&=0.5+0.5\tanh\frac{R-\sqrt{\left(x-100\right)^2+y^2}}{\epsilon/2}.
	\end{aligned}
\end{equation}

\begin{table}
	\centering
	\caption{The relative errors of the phase-field variables in the problem of the static droplets.}
	\begin{tabular}{ccccccccc}
		\toprule
		System && Phase && \emph{Model A}, $n=1$ && \emph{Model A}, $n=2$ && \emph{Model B} \\
		\midrule
		Quaternary && Phase 1 && $2.5877\times10^{-3}$ && $2.5877\times10^{-3}$ && $2.5877\times10^{-3}$ \\
		&& Phase 2 && $2.5877\times10^{-3}$ && $2.5877\times10^{-3}$ && $2.5877\times10^{-3}$ \\
		&& Phase 3 && $2.5876\times10^{-3}$ && $2.5876\times10^{-3}$ && $2.5876\times10^{-3}$ \\
		&& Phase 4 && $3.7739\times10^{-4}$ && $3.7739\times10^{-4}$ && $3.7739\times10^{-4}$ \\
		%\midrule
		Ternary && Phase 1 && $2.5877\times10^{-3}$ && $2.5877\times10^{-3}$ && $2.5877\times10^{-3}$ \\
		&& Phase 2 && $2.5877\times10^{-3}$ && $2.5877\times10^{-3}$ && $2.5877\times10^{-3}$ \\
		&& Phase 4 && $2.3993\times10^{-4}$ && $2.3993\times10^{-4}$ && $2.3993\times10^{-4}$ \\
		%\midrule
		Binary && Phase 1 && $2.5876\times10^{-3}$ && $2.5876\times10^{-3}$ && $2.5876\times10^{-3}$ \\
		&& Phase 4 && $1.1465\times10^{-4}$ && $1.1465\times10^{-4}$ && $1.1465\times10^{-4}$ \\
		\bottomrule
	\end{tabular}
	\label{table-droplets}
\end{table}

In the simulations, some parameters are set as $\rho_1:\rho_2:\rho_3:\rho_4=1000:500:100:1$, $\mu_1:\mu_2:\mu_3:\mu_4=100:50:10:0.1$, $\epsilon=5\Delta x$, $\Delta x=\Delta t=1$. The simple case of $\sigma_{pq}=0.01\,(p\neq q)$ is applied to test the accuracy of the present method and to probe the spurious velocities. The profiles of phase-field variables along the centerline are plotted in Fig. \ref{fig-droplets4} and the relative errors are listed in Table \ref{table-droplets}, where a good agreement between the numerical results and analytical solutions can be observed. One can also find that for this simple problem, two AC models show the same results for all cases. In addition, the consistency property of reduction is also tested through initializing the phase 3 (or phase 2 and phase 3) to be zero in the whole domain, and the comparison of numerical results and analytical solutions is shown in Fig. \ref{fig-droplets2}. From this figure, one can find that not only the numerical results are in agreement with the analytical solutions, but also the consistency of reduction is well preserved. Furthermore, we also report the histograms of the logarithm of the velocity magnitude ($p[\log_{10}(|\mathbf{u}|)]$) in Fig. \ref{fig-droplets-p}, where the magnitudes of spurious velocities of different systems are all of the order of $10^{-7}$, and there is no difference between \emph{Model A} and \emph{Model B}. 

\subsection{The spreading of a compound droplet on a solid wall}\label{wetting}
\begin{figure}
	\centering
	\includegraphics[width=2.5in]{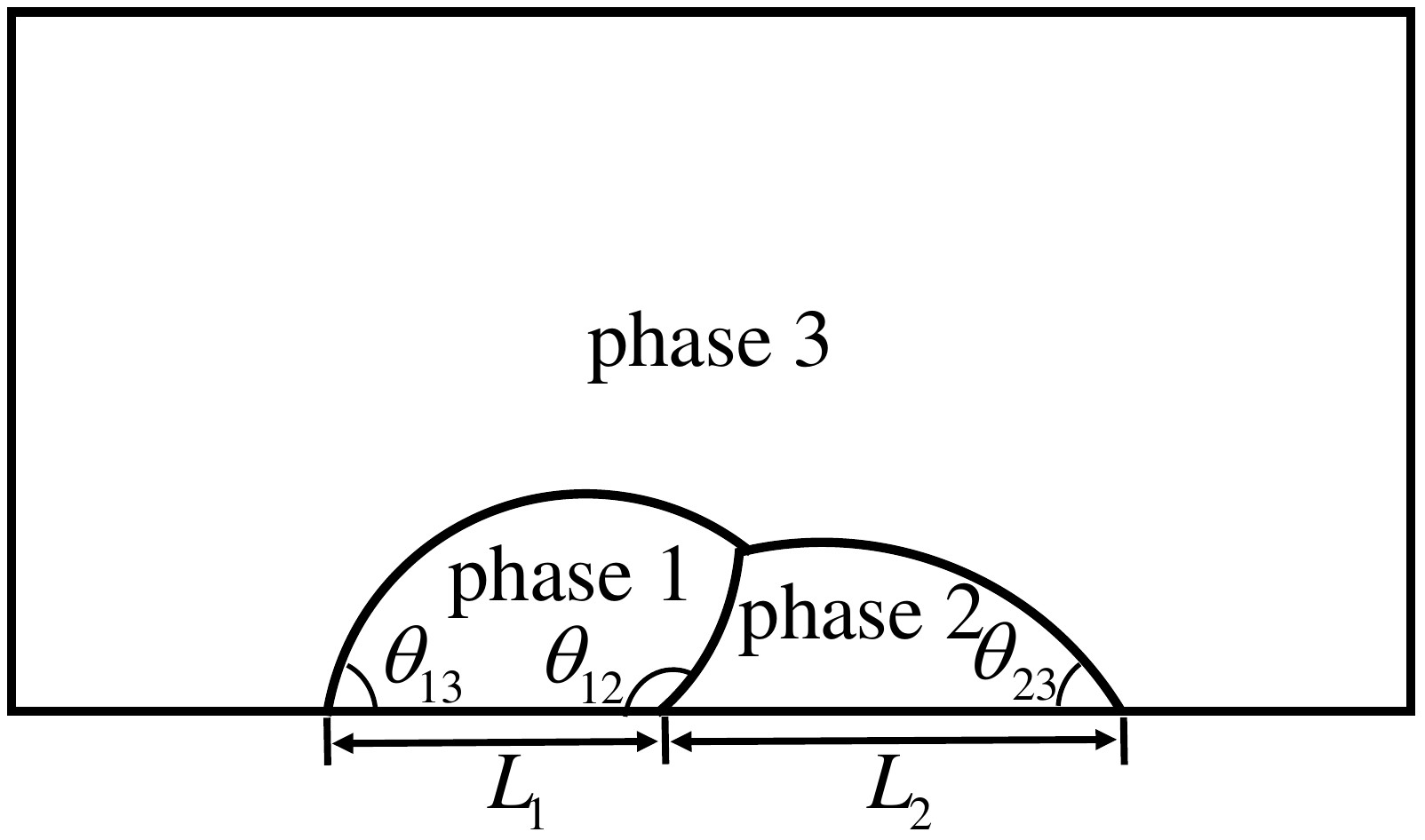}
	\caption{The configuration of the spreading of a compound droplet on a solid wall.}
	\label{fig-wet}
\end{figure}

The second problem is a compound droplet spreading on a solid wall, which is used to validate the wetting boundary condition for $N$-phase flows. In the computational domain $[-150,150]\times[0,150]$, a semicircular compound droplet consisting of phases 1 and 2 with radius $R=60$ is placed on the bottom wall. The periodic boundary condition is used in the horizontal direction, while the wetting and no-flux boundary conditions are imposed at the bottom and top boundaries, respectively. Figure \ref{fig-wet} is the schematic of this problem, and the phase-field variables are initialized by
\begin{equation}
	\begin{aligned}
		\phi_1\left(x,y\right)&=\left(0.5+0.5\tanh\frac{R-\sqrt{x^2+y^2}}{\epsilon/2}\right)\left(0.5-0.5\tanh\frac{2x}{\epsilon}\right),\\
		\phi_2\left(x,y\right)&=\left(0.5+0.5\tanh\frac{R-\sqrt{x^2+y^2}}{\epsilon/2}\right)\left(0.5+0.5\tanh\frac{2x}{\epsilon}\right).
	\end{aligned}
\end{equation}

In this work, the following wetting boundary condition is adopted \cite{Dong2017JCP,Liang2019AMM},
\begin{equation}
	\mathbf{n}_w\cdot\nabla\phi_p=\sum_{q\neq p}\frac{4}{\epsilon}\phi_p\phi_q\cos\theta_{pq},\quad\text{for }1\leq p\neq q\leq N,
\end{equation}
where $\mathbf{n}_w$ is the unit vector normal to the solid surface, $\theta_{pq}$ is the pairwise contact angle between the solid wall and the fluid interface formed by phases $p$ and $q$, measured on the side of phase $p$. Then we have $\cos\theta_{pq}=-\cos\theta_{qp}$, and the contact angle $\theta_{pq}$ can be expressed in terms of $N-1$ independent angles $\theta_{pN}$ with the Young's relation,
\begin{equation}
	\cos\theta_{pq}=\frac{\sigma_{pN}}{\sigma_{pq}}\cos\theta_{pN}-\frac{\sigma_{qN}}{\sigma_{pq}}\cos\theta_{qN},\quad\text{for }1\leq p\neq q\leq N-1,
\end{equation}
where $\theta_{pN}$ is the contact angle between the solid wall and the fluid interface formed by phases $p$ and $N$. 

\begin{figure}
	\centering
	\subfigure[$\theta_{23}=30^\circ$]{
		\begin{minipage}{0.48\linewidth}
			\centering
			\includegraphics[width=3in]{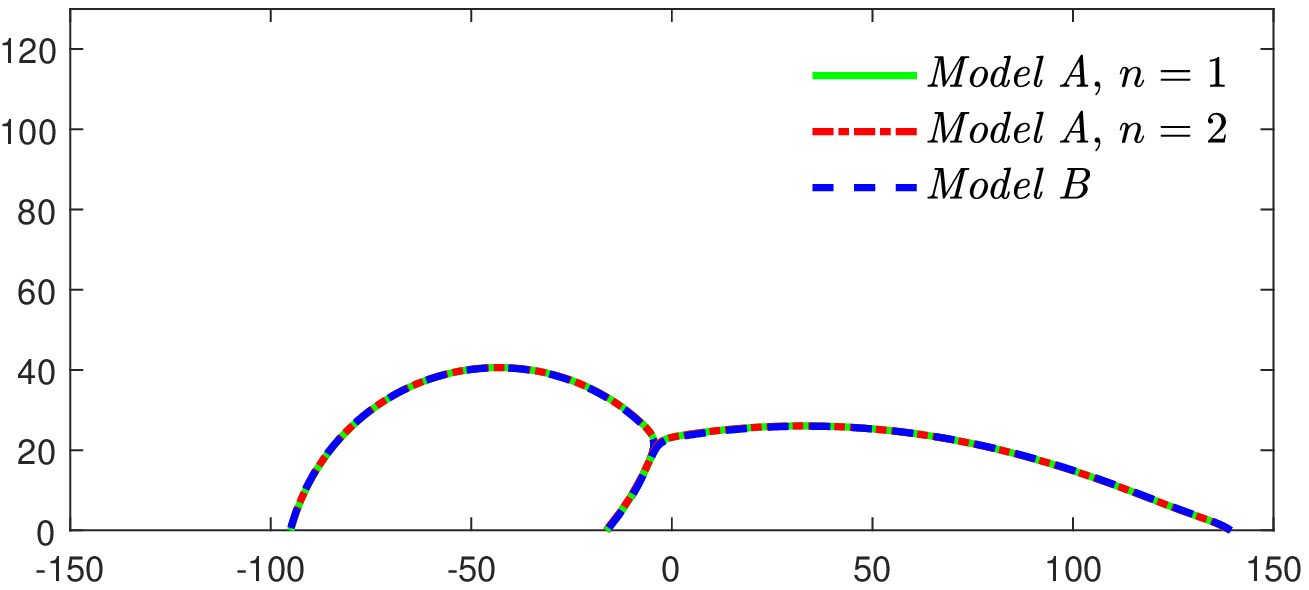}
	\end{minipage}}	
	\subfigure[$\theta_{23}=60^\circ$]{
		\begin{minipage}{0.48\linewidth}
			\centering
			\includegraphics[width=3in]{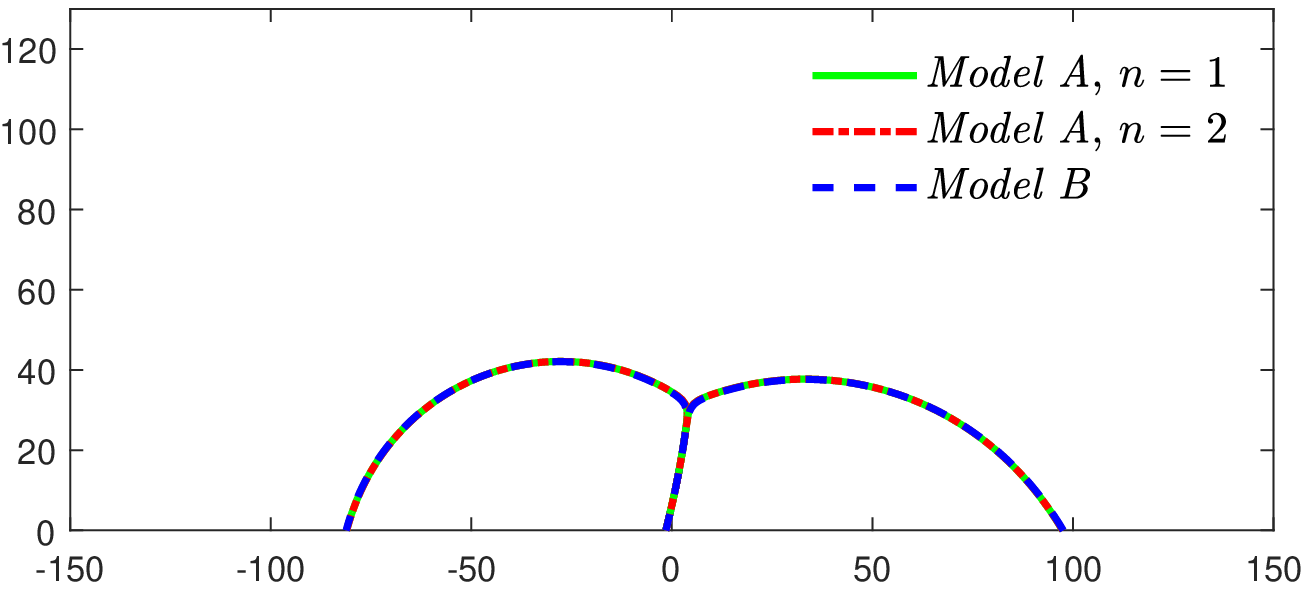}
	\end{minipage}}	
	\subfigure[$\theta_{23}=90^\circ$]{
		\begin{minipage}{0.48\linewidth}
			\centering
			\includegraphics[width=3in]{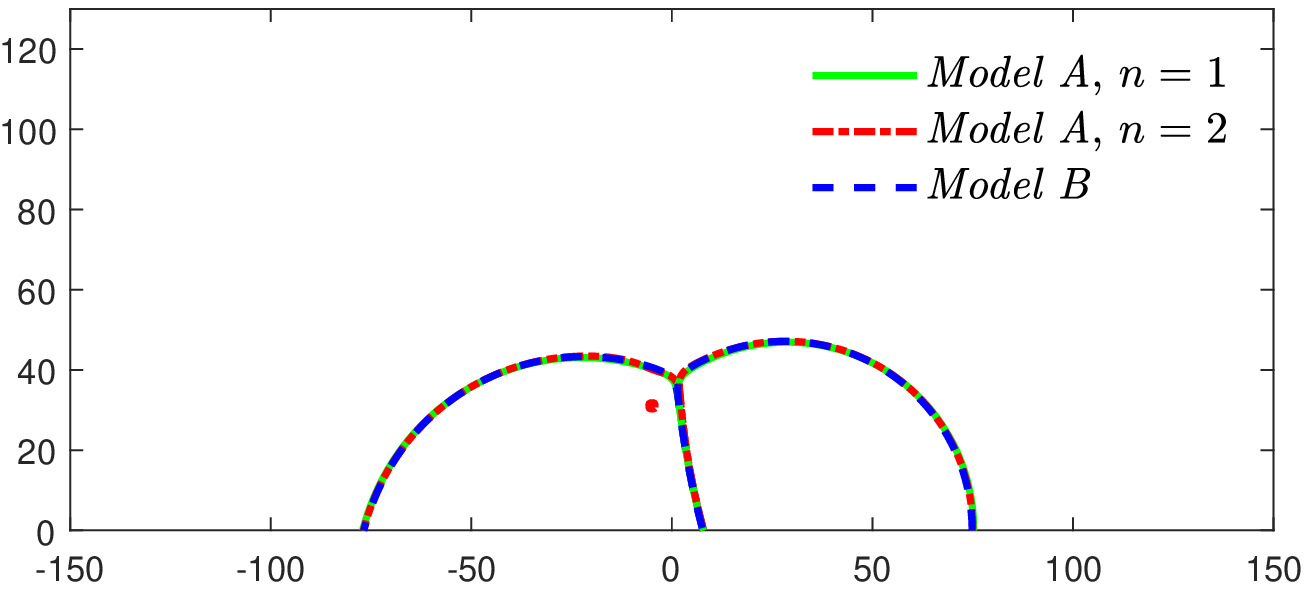}
	\end{minipage}}	
	\subfigure[$\theta_{23}=120^\circ$]{
		\begin{minipage}{0.48\linewidth}
			\centering
			\includegraphics[width=3in]{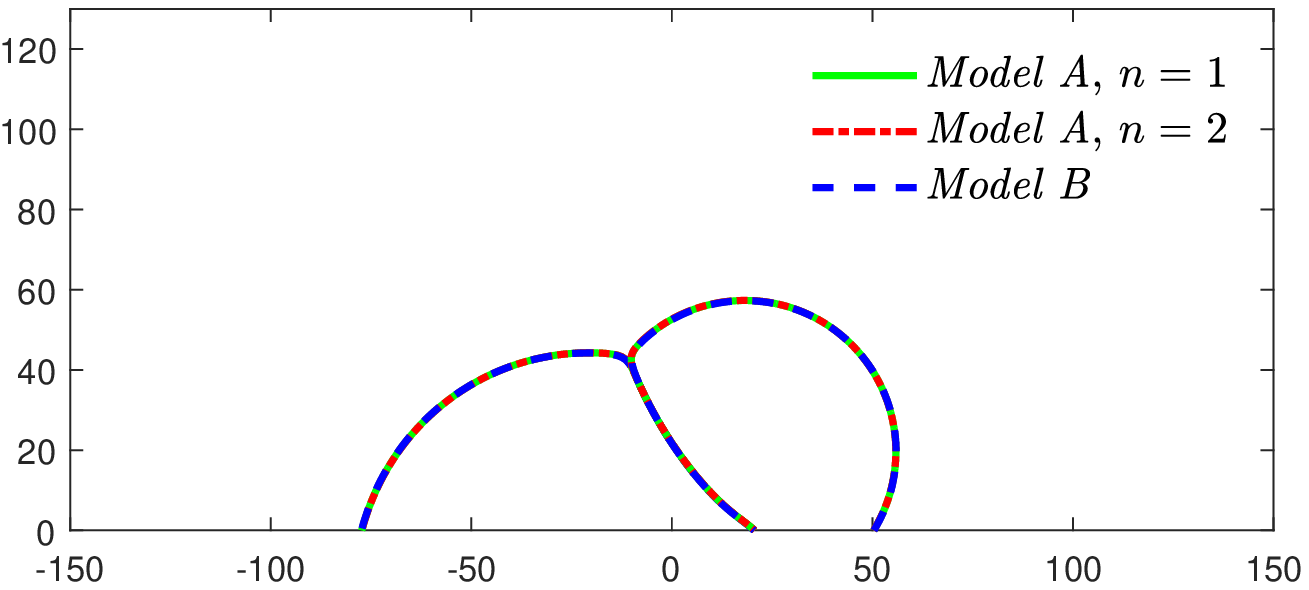}
	\end{minipage}}	
	\caption{The equilibrium distributions of the three fluids with contact angle $\theta_{13}=75^\circ$ and different values of $\theta_{23}$.}
	\label{fig-wetting}
\end{figure}

In this section, the material properties and lattice settings are the same as the case of ternary system in Sec. \ref{droplets}. We conduct some simulations in the case of $\theta_{13}=75^\circ$, and plot the finial shapes of the droplet in Fig. \ref{fig-wetting}. From this figure, one can observe that the wetted length of fluid 2 ($L_2$) decreases a lot with the increase of $\theta_{23}$, while that of fluid 1 ($L_1$) increases less. There are also small differences between two AC models, but in the case of $\theta_{23}=90^\circ$ [see Fig. \ref{fig-wetting}(c)], an unphysical phenomenon can be found in \emph{Model A} with $n=2$, which states that \emph{Model A} is more inaccurate. To quantitatively show the accuracy of present LB method, we also measure the numerical wetted lengths of fluid 1 and fluid 2, and compared them with the analytical solutions in Table \ref{table-wetting}, from which one can find that the numerical results of two AC models are close to the analytical solutions, and their relative errors are in the same orders. 

\begin{table}
	\centering
	\caption{The equilibrium wetted lengths $L_1$ and $L_2$ (normalized by the initial radius $R$) in the spreading of a compound droplet with $\theta_{13}=75^\circ$.}
	\begin{tabular}{lcccccccccc}
		\toprule
		 &&&& \multicolumn{3}{c}{Numerical} && \multicolumn{3}{c}{Relative error}\\
		\cline{5-7}\cline{9-11}
		Case & Length & Analytical && \emph{Model A}, & \emph{Model A}, & \emph{Model B} &&  \emph{Model A}, & \emph{Model A}, & \emph{Model B} \\
		& & && $n=1$ & $n=2$ & && $n=1$ & $n=2$ & \\
		\midrule
		$\theta_{23}=30^\circ$ & $L_1$ & 1.3482 && 1.3126 & 1.3123 & 1.3170 && 2.64\% & 2.66\% & 2.31\% \\
		& $L_2$ & 2.4862 && 2.5928 & 2.5921 & 2.5928 && 4.29\% & 4.26\% & 4.29\% \\ 
		$\theta_{23}=60^\circ$ & $L_1$ & 1.3563 && 1.3266 & 1.3264 & 1.3286 && 2.19\% & 2.20\% & 2.04\% \\
		& $L_2$ & 1.6666 && 1.6491 & 1.6492 & 1.6512 && 1.05\% & 1.04\% & 0.92\% \\
		$\theta_{23}=90^\circ$ & $L_1$ & 1.4328 && 1.4161 & 1.4139 & 1.4092 && 1.32\% & 1.32\% & 1.65\% \\
		& $L_2$ & 1.1245 && 1.1233 & 1.1185 & 1.1183 && 0.11\% & 0.53\% & 0.55\% \\
		$\theta_{23}=120^\circ$ & $L_1$ & 1.6213 && 1.6320 & 1.6321 & 1.6313 && 0.66\% & 0.67\% & 0.62\% \\
		& $L_2$ & 0.5334 && 0.4959 & 0.4951 & 0.4988 && 7.03\% & 7.18\% & 6.49\% \\
		\bottomrule
	\end{tabular}
	\label{table-wetting}
\end{table}

\subsection{The spreading of a liquid lens}\label{lens}
\begin{figure}
	\centering
	\includegraphics[width=3.0in]{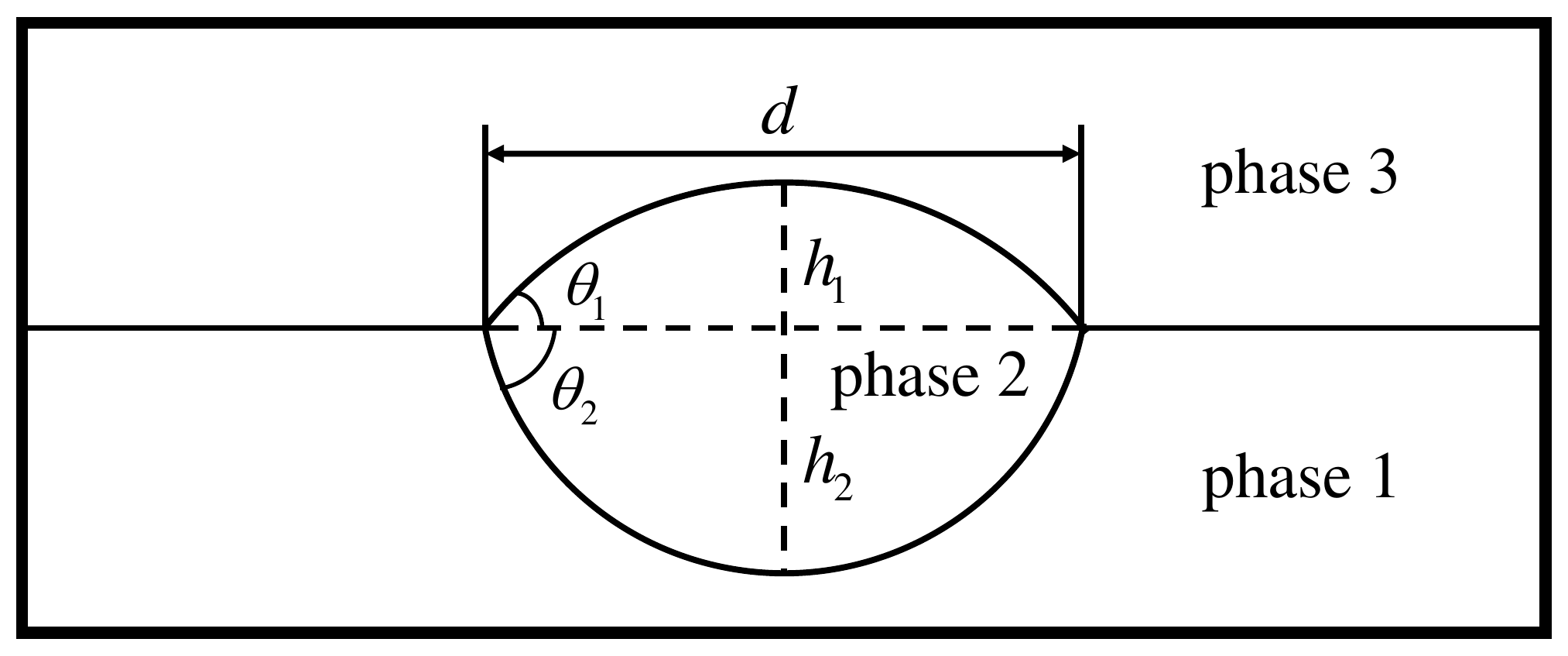}
	\caption{The configuration of the spreading of a liquid lens.}
	\label{fig-lens}
\end{figure}

We now focus on the spreading of a liquid lens to investigate the equilibrium states of the multiphase flows. This benchmark problem has been widely used to test the ternary-fluid models \cite{Boyer2006MMAN,Kim2007CMAME,Liang2016PRE,Abadi2018PRE,Yu2019PF,Dong2018JCP,Zheng2020PREa,Yuan2022PF} because of the available analytical (or asymptotic) solutions. In our simulations, the computational domain is $[-0.04,0.04]\times[-0.016,0.016]$ with periodic boundary conditions at the left and right boundaries, while no-flux walls at the top and the bottom. The schematic of the problem is shown in Fig. \ref{fig-lens} where a circular droplet of phase 2 with radius $R=8\times10^{-3}$ is initially placed at the interface of a binary fluid layer. The phase-field variables are initialized by  
\begin{equation}
	\begin{aligned}
		\phi_2\left(x,y\right)&=0.5+0.5\tanh\frac{R-\sqrt{x^2+y^2}}{\epsilon/2},\\
		\phi_1\left(x,y\right)&=\left[1-\phi_2\left(x,y\right)\right]\left(0.5+0.5\tanh\frac{-2y}{\epsilon}\right).%,\\
		%\phi_3\left(x,y\right)&=1-\phi_1\left(x,y\right)-\phi_2\left(x,y\right).
	\end{aligned}
\end{equation} 

\begin{figure}
	\centering
	\subfigure[$\sigma_{12}=0.0050$, $\sigma_{23}=0.0087$]{
		\begin{minipage}{0.48\linewidth}
			\centering
			\includegraphics[width=3in]{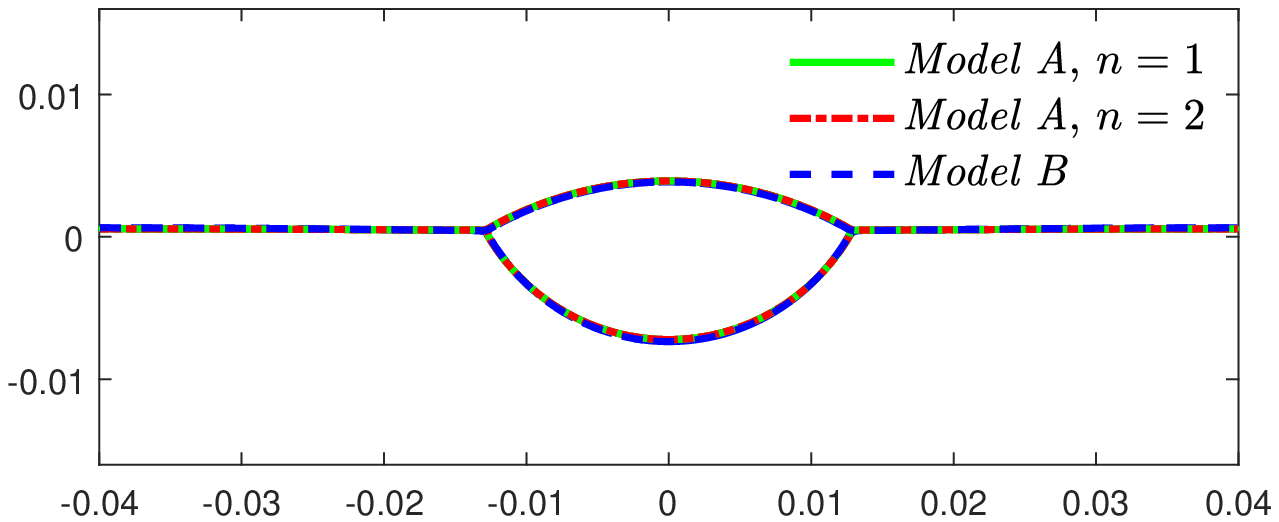}
	\end{minipage}}
	\subfigure[$\sigma_{12}=0.0058$, $\sigma_{23}=0.0115$]{
		\begin{minipage}{0.48\linewidth}
			\centering
			\includegraphics[width=3in]{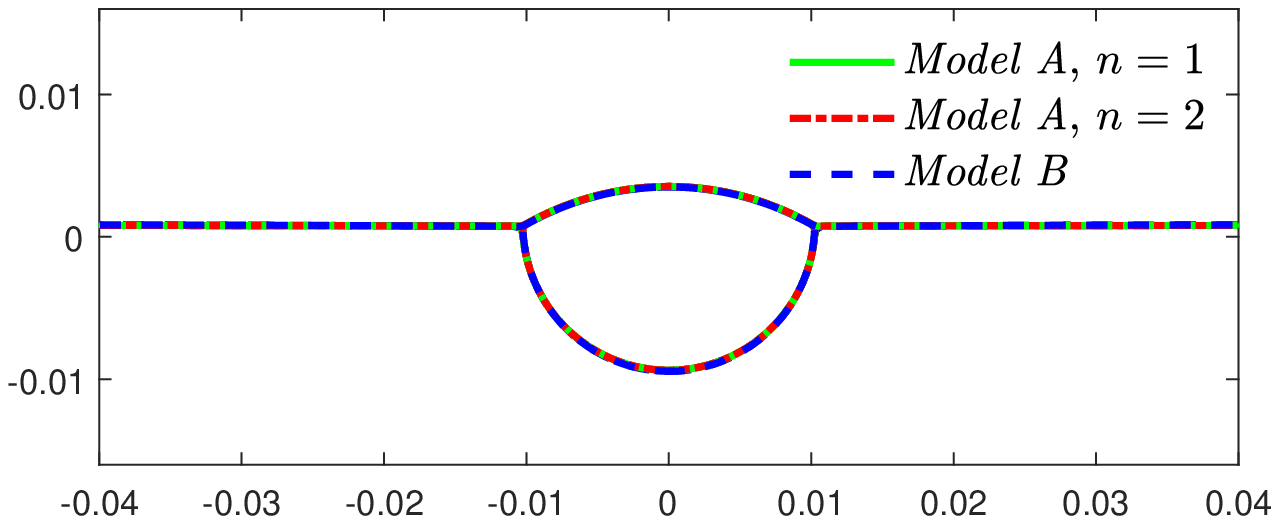}
	\end{minipage}}
	\subfigure[$\sigma_{12}=0.0100$, $\sigma_{23}=0.0100$]{
		\begin{minipage}{0.48\linewidth}
			\centering
			\includegraphics[width=3in]{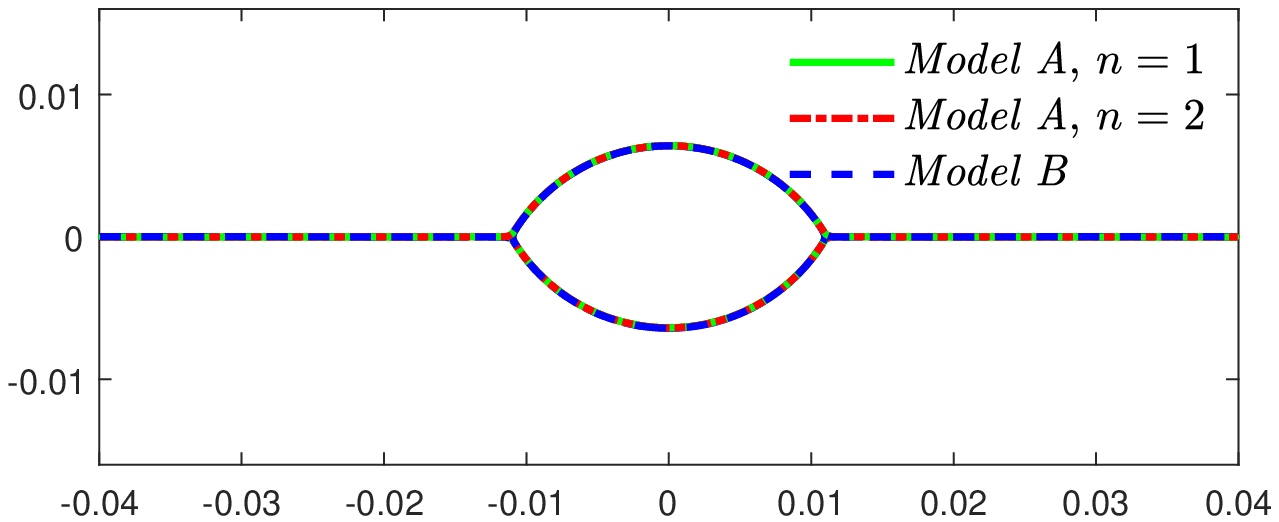}
	\end{minipage}}
	\subfigure[$\sigma_{12}=0.0173$, $\sigma_{23}=0.0200$]{
		\begin{minipage}{0.48\linewidth}
			\centering
			\includegraphics[width=3in]{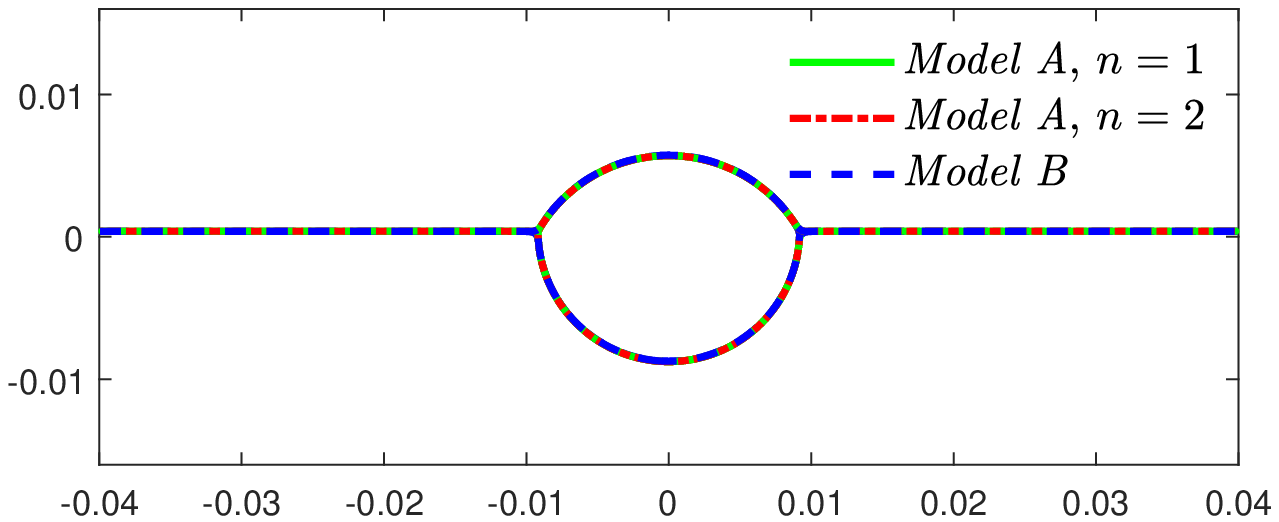}
	\end{minipage}}
	\caption{The equilibrium shapes of the liquid lens with different surface tension coefficient ratios.}
	\label{fig-lens-G0}
\end{figure}

In the absence of the gravity, the droplet of phase 2 will form a lens at the equilibrium state dominated by the surface tension, and the equilibrium contact angles can be obtained by the Neumann's law \cite{Rowlinson1982},
\begin{equation}
	\cos\theta_1=\frac{\sigma_{13}^2+\sigma_{23}^2-\sigma_{12}^2}{2\sigma_{13}\sigma_{23}},\quad \cos\theta_2=\frac{\sigma_{12}^2+\sigma_{13}^2-\sigma_{23}^2}{2\sigma_{12}\sigma_{13}}.
\end{equation}
According to above equation and geometrical relation, the distance between the triple junction points ($d$) and the heights of the lens ($h_1$, $h_2$) can be calculated by
\begin{equation}\label{eq-height}
	d=2\sqrt{\frac{S}{\sum_{i=1}^2\frac{1}{\sin\theta_i}\left(\frac{\theta_i}{\sin\theta_i}-\cos\theta_i\right)}},\quad h_i=\frac{d}{2}\frac{1-\cos\theta_i}{\sin\theta_i},\quad i=1,2,
\end{equation}
where $S$ is the area of the lens. In this case, we set some parameters as $\rho_1:\rho_2:\rho_3=1000:500:1$, $\mu_1:\mu_2:\mu_3=0.1:0.05:0.0001$, $\Delta x=1.25\times10^{-4}$, $\Delta t=\Delta x/20$, and conduct some simulations with different surface tension coefficient ratios: (a) $\sigma_{12}=0.005$, $\sigma_{23}=0.0087$; (b) $\sigma_{12}=0.0058$, $\sigma_{23}=0.0115$; (c) $\sigma_{12}=0.01$, $\sigma_{23}=0.01$; (d) $\sigma_{12}=0.0173$, $\sigma_{23}=0.02$, and $\sigma_{13}=0.01$ in all cases. The equilibrium states of the lens for cases (a)-(d) are shown in Fig. \ref{fig-lens-G0}, where the geometrical sizes of lens are related to different surface tension coefficient ratios, and the configurations of the liquid lens are in good agreement with those reported in Ref. \cite{Yu2019PF}. To give a quantitative test, we also measure the length ($d$) and heights ($h_1$, $h_2$) of the droplet lens and compare them with the analytical solutions. Table \ref{table-lens} lists the relative errors of above characteristic quantities, they are less than 3\% and some little differences are shown among two AC models.

\begin{figure}
	\centering
	\subfigure[$|\mathbf{g}|=0$]{
		\begin{minipage}{0.48\linewidth}
			\centering
			\includegraphics[width=3in]{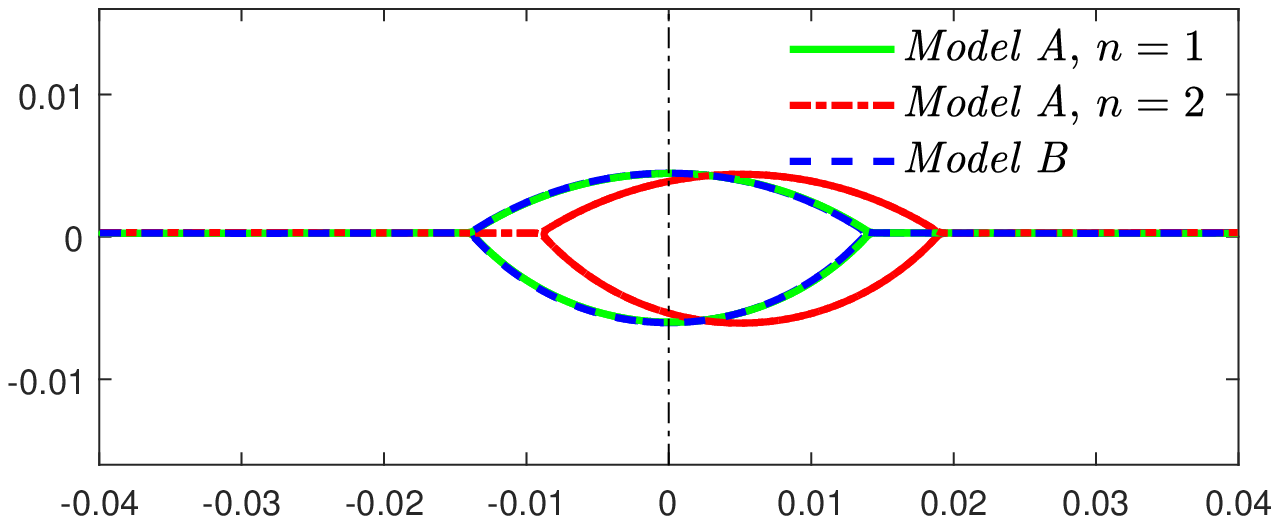}
	\end{minipage}}
	\subfigure[$|\mathbf{g}|=2$]{
		\begin{minipage}{0.48\linewidth}
			\centering
			\includegraphics[width=3in]{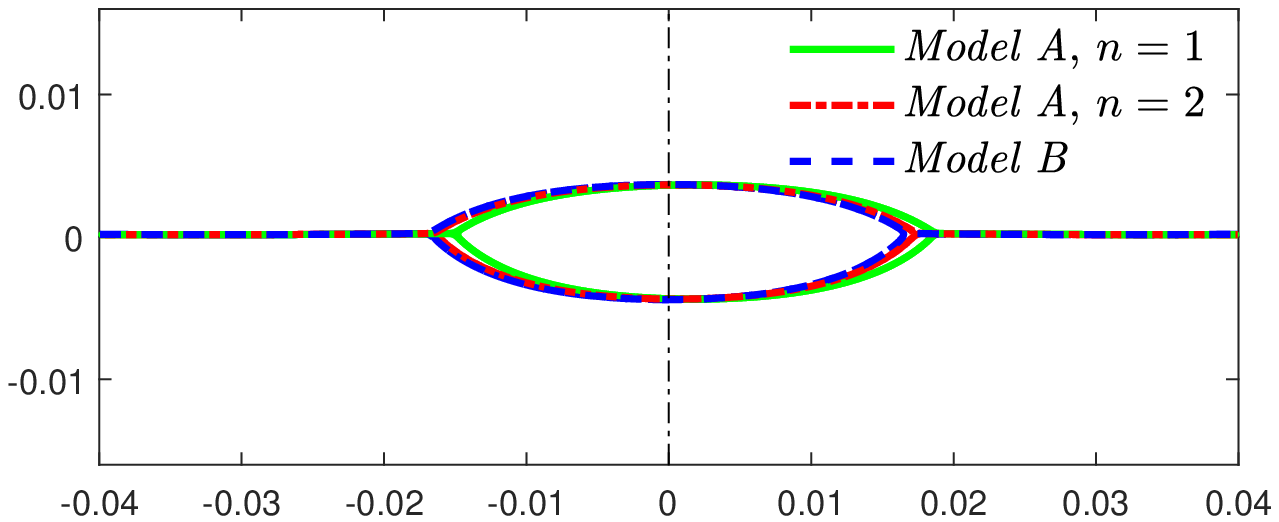}
	\end{minipage}}
	\caption{A comparison of the equilibrium shapes of the oil lens at $t=8$\si{s}.}
	\label{fig-lens-G0com}
\end{figure}

\begin{figure}
	\centering
	\subfigure[$|\mathbf{g}|=2$]{
	\begin{minipage}{0.48\linewidth}
			\centering
			\includegraphics[width=2.0in]{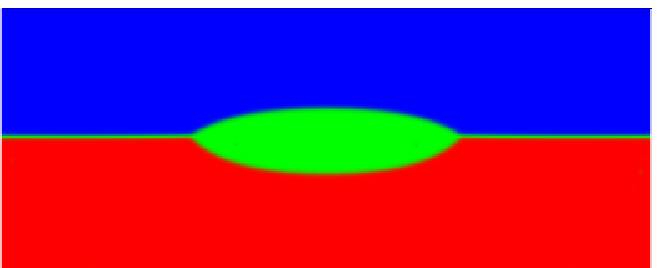}
	\end{minipage}}
	\subfigure[$|\mathbf{g}|=5$]{
		\begin{minipage}{0.48\linewidth}
			\centering
			\includegraphics[width=2.0in]{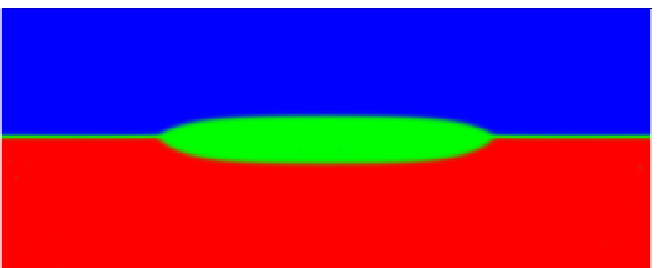}
	\end{minipage}}
	\subfigure[$|\mathbf{g}|=7.5$]{
		\begin{minipage}{0.48\linewidth}
			\centering
			\includegraphics[width=2.0in]{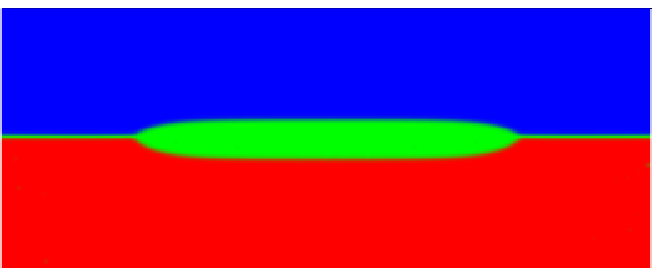}
	\end{minipage}}
	\subfigure[$|\mathbf{g}|=9.8$]{
		\begin{minipage}{0.48\linewidth}
			\centering
			\includegraphics[width=2.0in]{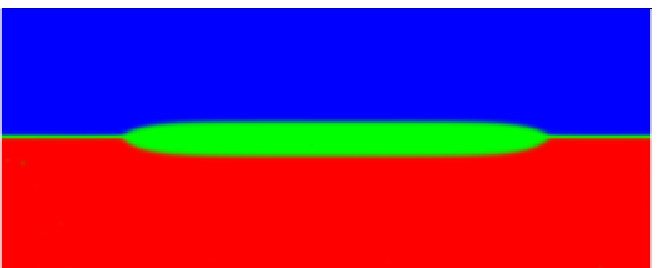}
	\end{minipage}}
	\caption{The equilibrium shapes of the oil lens at different magnitudes of the gravity (\emph{Model B}).}
	\label{fig-lens-G}
\end{figure}

\begin{figure}
	\centering
	\includegraphics[width=3.5in]{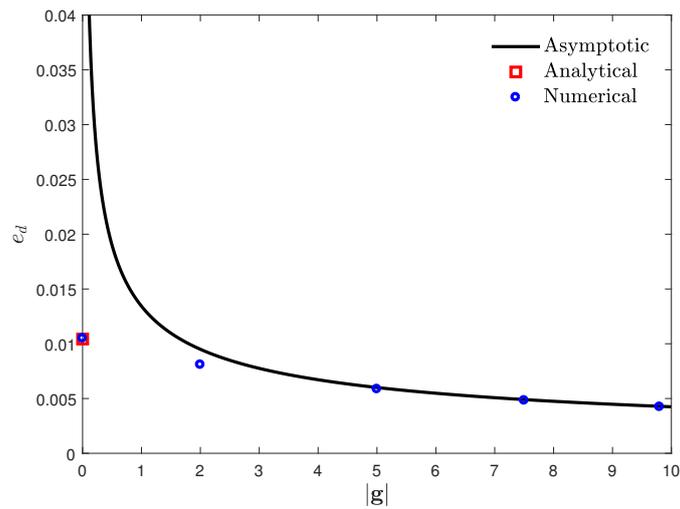}
	\caption{A comparison of the asymptotic, analytical, and the current numerical results (\emph{Model B}) of equilibrium thickness of the lens with different magnitudes of the gravity.}
	\label{fig-lens-com}
\end{figure}

In the presence of gravity, we consider an actual ternary system consisting of an oil droplet sandwiched between the air and water layers, and the material properties are shown in Table \ref{table-material}. In this realistic case, however, as shown by the preliminary results in Fig. \ref{fig-lens-G0com}, the oil lens of \emph{Model A} is no longer in a static state in most cases even at $|\mathbf{g}|=0$, which indicates that \emph{Model A} is inaccurate for realistic multiphase flows. For this reason, only \emph{Model B} is used in the following simulations of oil lens. We consider four different magnitudes of the gravity, $|\mathbf{g}|=2,5,7.5,9.8$, and show the final shapes of the oil droplet in Fig. \ref{fig-lens-G}. From this figure, one can observe that the oil droplet is horizontally elongated and vertically compressed with the increase of gravity. When the gravity is dominate, the asymptotic result from Langmuir-de Gennes \cite{Langmuir1933JCP,Gennes2004} provides the thickness of the oil lens,
\begin{equation}\label{eq-thickness}
	e_d=\sqrt{\frac{2\left(\sigma_{12}+\sigma_{23}-\sigma_{13}\right)\rho_1}{\left(\rho_1-\rho_2\right)\rho_2|\mathbf{g}|}}.
\end{equation}
We also measure the thicknesses of the oil lens under different magnitudes of gravity, and compare them with above analytical solution (\ref{eq-height}) and asymptotic solution (\ref{eq-thickness}) in Fig. \ref{fig-lens-com}. Actually, the Langmuir-de Gennes theory only holds with the large magnitudes of gravity, and in this limit, our numerical results indeed match the theory quite well.

\begin{table}
	\centering
	\caption{The equilibrium length and heights (normalized by the initial radius $R$) of the liquid lens in the absence of gravity in different cases.}
	\begin{tabular}{ccccccccccccc}
		\toprule
		&& \multicolumn{3}{c}{Analytical solutions} && \multicolumn{3}{c}{Numerical results} && \multicolumn{3}{c}{Relative errors}\\
		\cline{3-5}\cline{7-9}\cline{11-13}
		Model & Case & $d$ & $h_1$ & $h_2$ && $d$ & $h_1$ & $h_2$ && $d$ & $h_1$ & $h_2$\\
		\midrule
		\emph{Model A}, & (a) & 1.6250 & 0.4354 & 0.9468 && 1.5876 & 0.4507 & 0.9416 && 2.30\% & 3.52\% & 0.56\% \\
		$n=1$& (b) & 1.2815 & 0.3468 & 1.2662 && 1.2815 & 0.3484 & 1.2651 && 0.00\% & 0.47\% & 0.08\% \\
		& (c) & 1.3850 & 0.7996 & 0.7996 && 1.3713 & 0.8007 & 0.7981 && 0.99\% & 0.14\% & 0.19\% \\
		& (d) & 1.1461 & 0.6601 & 1.1485 && 1.1477 & 0.6658 & 1.1418 && 0.14\% & 0.86\% & 0.58\% \\
		\emph{Model A}, & (a) & 1.6250 & 0.4354 & 0.9468 && 1.5896 & 0.4511 & 0.9406 && 2.18\% & 3.60\% & 0.66\% \\
		$n=2$& (b) & 1.2815 & 0.3468 & 1.2662 && 1.2815 & 0.3486 & 1.2650 && 0.03\% & 0.52\% & 0.09\% \\
		& (c) & 1.3850 & 0.7996 & 0.7996 && 1.3710 & 0.8002 & 0.7988 && 1.01\% & 0.07\% & 0.10\% \\
		& (d) & 1.1461 & 0.6601 & 1.1485 && 1.1473 & 0.6660 & 1.1420 && 0.10\% & 0.88\% & 0.56\% \\
		\emph{Model B} & (a) & 1.6250 & 0.4354 & 0.9468 && 1.5746 & 0.4508 & 0.9506 && 3.10\% & 3.54\% & 0.40\% \\
		& (b) & 1.2815 & 0.3468 & 1.2662 && 1.2782 & 0.3472 & 1.2703 && 0.26\% & 0.10\% & 0.33\% \\
		& (c) & 1.3850 & 0.7996 & 0.7996 && 1.3682 & 0.8058 & 0.7947 && 1.21\% & 0.78\% & 0.62\% \\
		& (d) & 1.1461 & 0.6601 & 1.1485 && 1.1473 & 0.6618 & 1.1464 && 0.11\% & 0.25\% & 0.18\% \\
		\bottomrule
	\end{tabular}
	\label{table-lens}
\end{table}
\begin{table}
	\centering
	\caption{Material properties of the water, oil, and air.}
	\begin{tabular}{lllllll}
		\toprule
		Density [\si{kg/m^3}] && Water: $998.207$ && Oil: $557$ && Air: $1.204$\\
		Dynamic viscosity [\si{kg/(m.s)}] && Water: $1.002\times10^{-3}$ && Oil: $9.15\times10^{-2}$ && Air: $1.78\times10^{-5}$ \\
		Surface tension [\si{kg/s^2}]  && Water/Oil: $0.04$ && Oil/Air: $0.055$ && Water/Air: $0.0728$ \\
		Magnitude of gravity [\si{m/s^2}] && 9.8 && && \\
		\bottomrule
	\end{tabular}
	\label{table-material}
\end{table}

\subsection{Rayleigh–Taylor instability}\label{RTI}
\begin{figure}
	\centering
	\subfigure[$T=0$]{
		\begin{minipage}{0.12\linewidth}
			\centering
			\includegraphics[width=.75in]{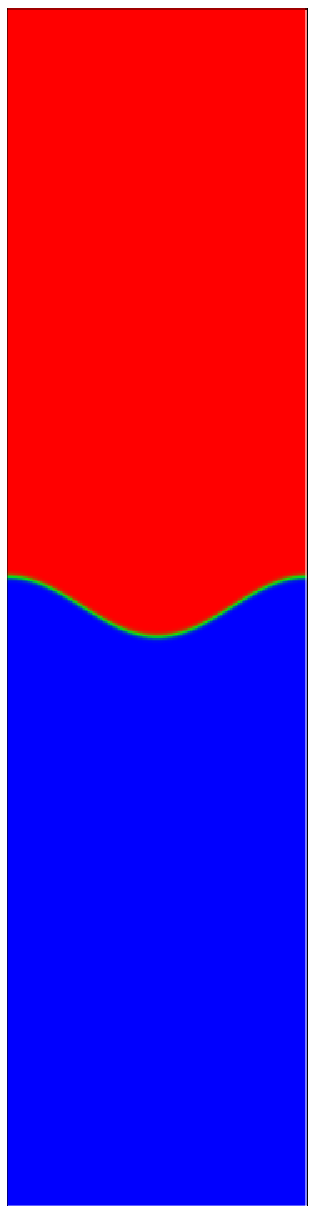}
	\end{minipage}}
	\subfigure[$T=0.5$]{
		\begin{minipage}{0.12\linewidth}
			\centering
			\includegraphics[width=.75in]{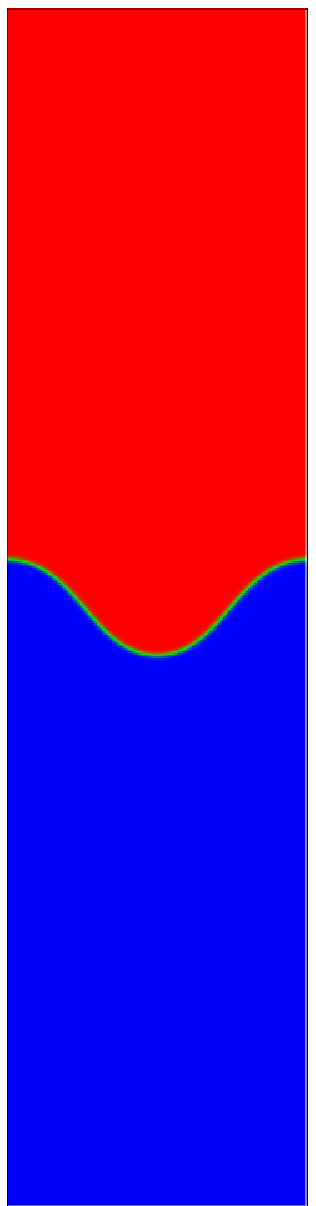}
	\end{minipage}}
	\subfigure[$T=1$]{
		\begin{minipage}{0.12\linewidth}
			\centering
			\includegraphics[width=.75in]{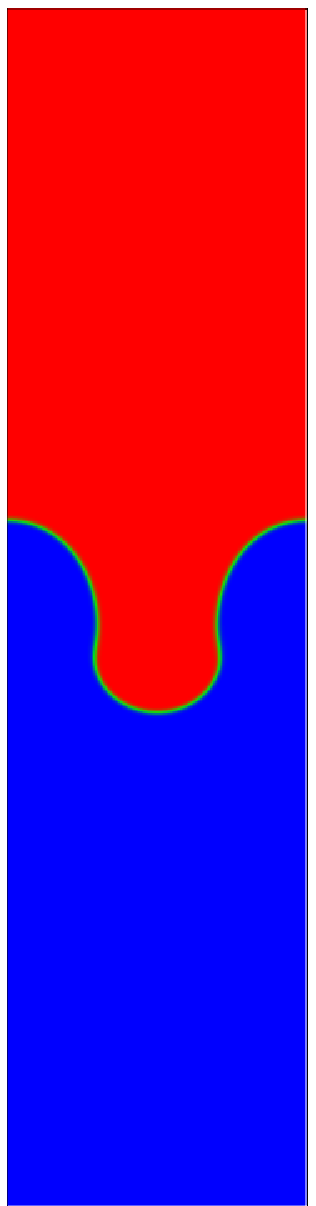}
	\end{minipage}}
	\subfigure[$T=1.5$]{
		\begin{minipage}{0.12\linewidth}
			\centering
			\includegraphics[width=.75in]{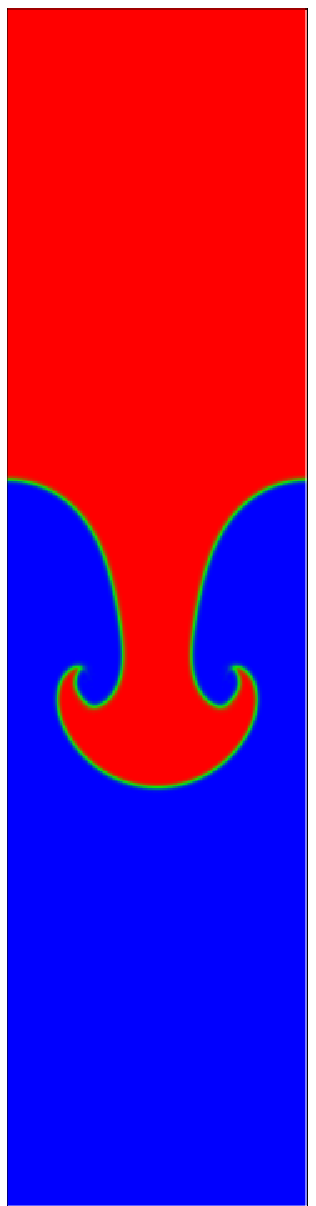}
	\end{minipage}}
	\subfigure[$T=2$]{
		\begin{minipage}{0.12\linewidth}
			\centering
			\includegraphics[width=.75in]{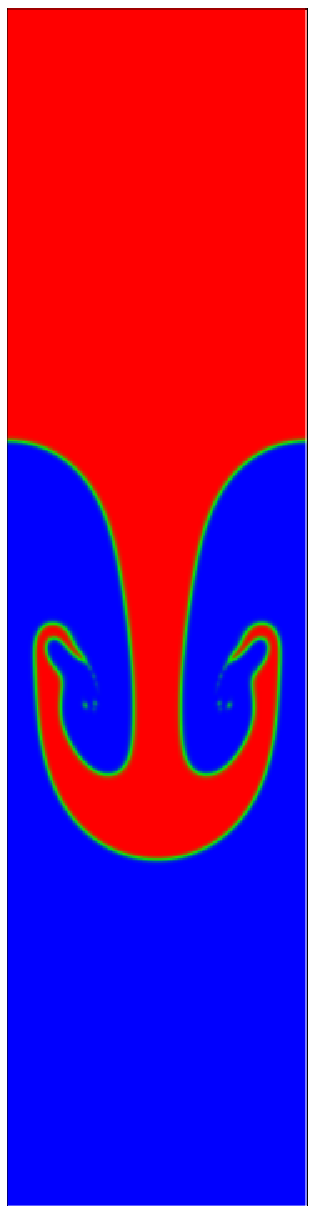}
	\end{minipage}}
	\subfigure[$T=2.5$]{
		\begin{minipage}{0.12\linewidth}
			\centering
			\includegraphics[width=.75in]{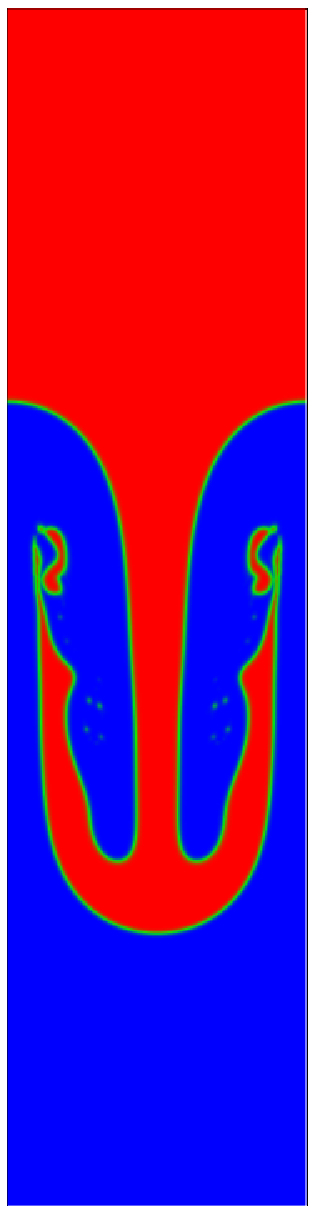}
	\end{minipage}}
	\subfigure[$T=3$]{
		\begin{minipage}{0.12\linewidth}
			\centering
			\includegraphics[width=.75in]{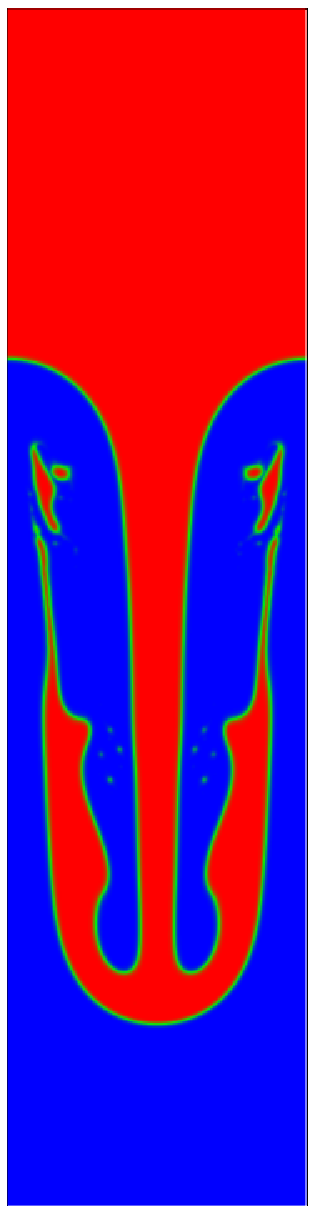}
	\end{minipage}}
	\caption{The snapshots of the RTI in the binary system.}
	\label{fig-RTI2}
\end{figure}
\begin{figure}
	\centering
	\subfigure[]{
		\begin{minipage}{0.48\linewidth}
			\centering
			\includegraphics[width=3.0in]{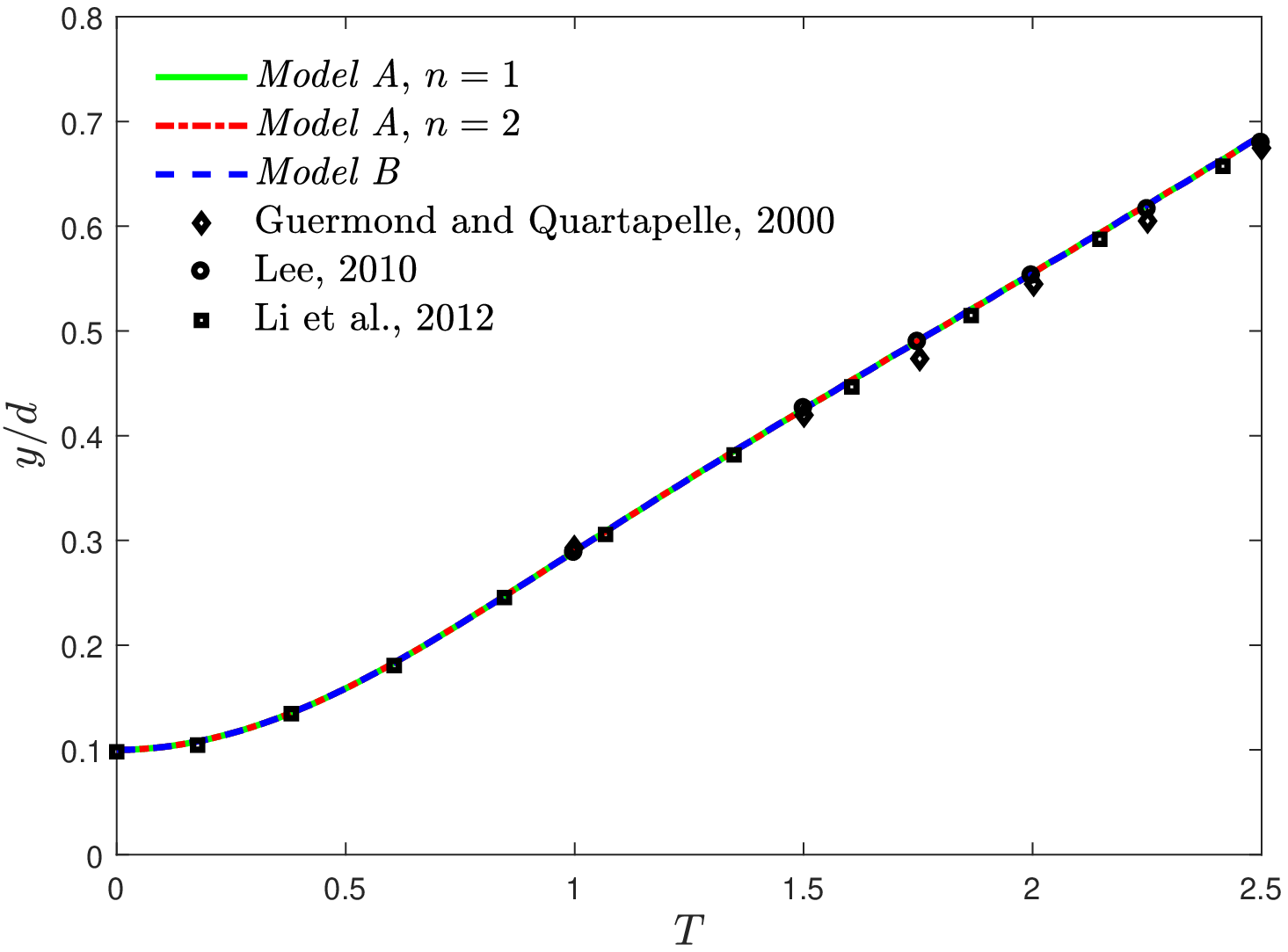}
	\end{minipage}}
	\subfigure[]{
		\begin{minipage}{0.48\linewidth}
			\centering
			\includegraphics[width=3.0in]{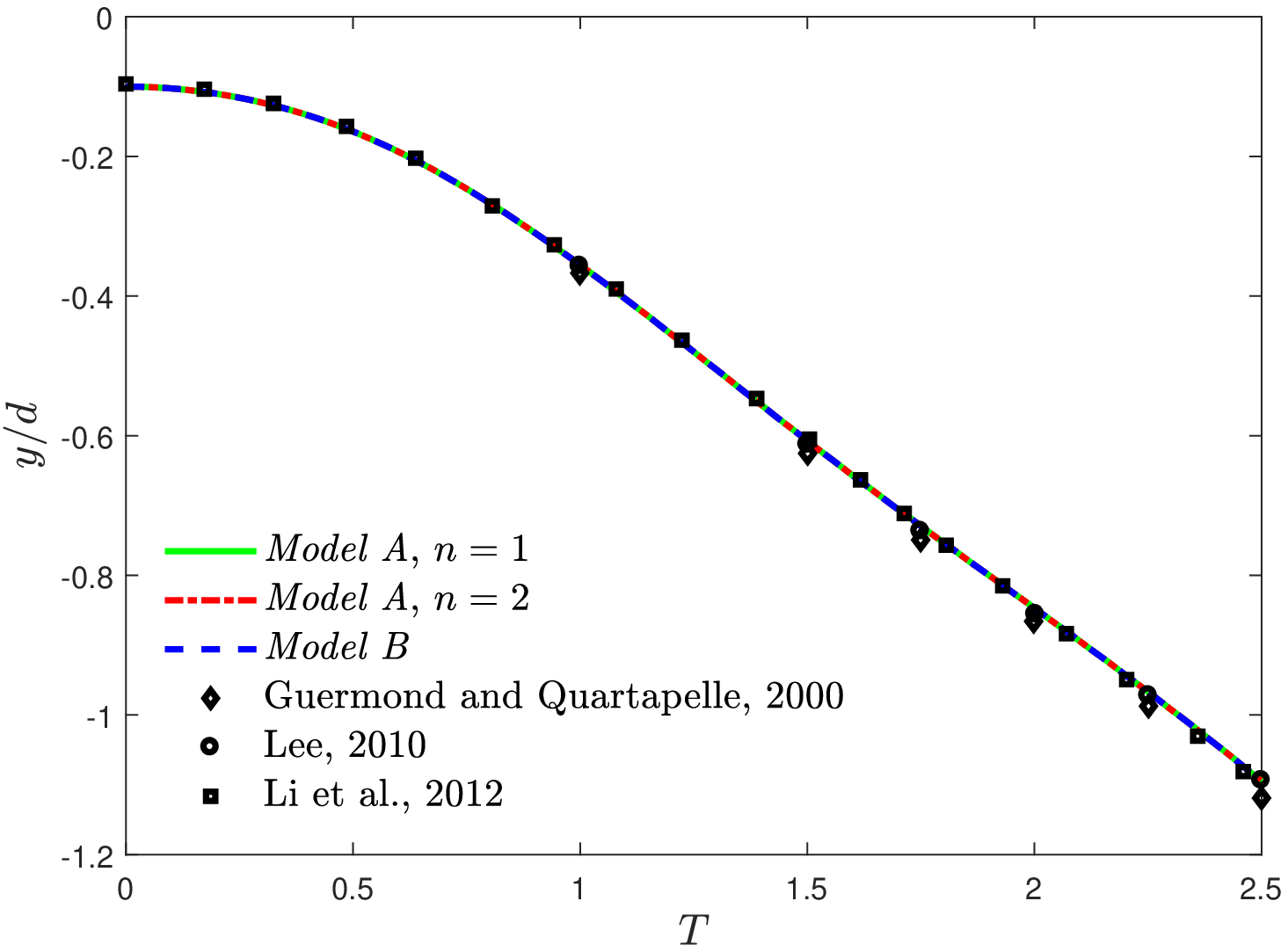}
	\end{minipage}}
	\caption{The normalized positions of the tips of the rising and falling fluids of the RTI in the binary system.}
	\label{fig-RTI2-com}
\end{figure}
The RTI is a classical example in multiphase flows \cite{Keskinen1981JGR,Zhou2017PR}, and it occurs when the interface is perturbed while a heavier fluid is sitting on the top of the light one in the gravitational field. Here, the unsteady ternary-fluid RTI is used to further compare the performances of two AC models for the problems with large deformation of the interfaces. The domain considered here is $[-d/2,d/2]\times[-2d,2d]$ with the periodic boundary condition in the horizontal direction and no-flux condition on the top and bottom boundaries. The RTI is usually depicted by the Atwood number $At=\left(\rho^*-1\right)/\left(\rho^*+1\right)$, Reynolds number $Re=d\rho_3\sqrt{gd}/\mu_3$, and capillary number $Ca=\mu_3\sqrt{gd}/\sigma$, where $\rho^*$ is the density ratio between the heaver fluid (phase 1) and the lighter fluid (phase 3), and $g$ is the magnitude of the gravitational acceleration. In the simulations, $\Delta x=\Delta t=1$, the grid size is $256\times1024$, and the dimensionless time is defined by $T=t/\sqrt{d/gAt}$. Other physical parameters are fixed as $At=0.5$, $Re=1000$, $Ca=0.28$, $\rho_1:\rho_2:\rho_3=3:2:1$, and $\mu_1:\mu_2:\mu_3=1$. 

We first consider the binary system by setting $\phi_2=0$ according to the consistency property of reduction, and the initial phase-field variables are given by
\begin{equation}
	\begin{aligned}
		\phi_1\left(x,y\right)&=0.5+0.5\tanh\frac{y-0.1d\cos\left(2\pi x/d\right)}{\epsilon/2},\\
		\phi_2\left(x,y\right)&=0.
	\end{aligned}
\end{equation}
The evolution of the interface of RTI in the binary-fluid case is shown in Fig. \ref{fig-RTI2}, where the present results are in good agreement with the available data in Refs. \cite{Zu2013PRE,Ren2016PRE,Liu2023PRE}. To give a quantitative comparison, the normalized positions of the top of the rising fluid and the bottom of the falling fluid are also plotted in Fig. \ref{fig-RTI2-com}. From this figure, one can find that there is an agreement among present results and previous works \cite{Guermond2000JCP,Lee2010CF,Li2012PRE}, and no obvious difference between two AC models.

\begin{figure}
	\centering
	\subfigure[$T=0$]{
		\begin{minipage}{0.12\linewidth}
			\centering
			\includegraphics[width=.75in]{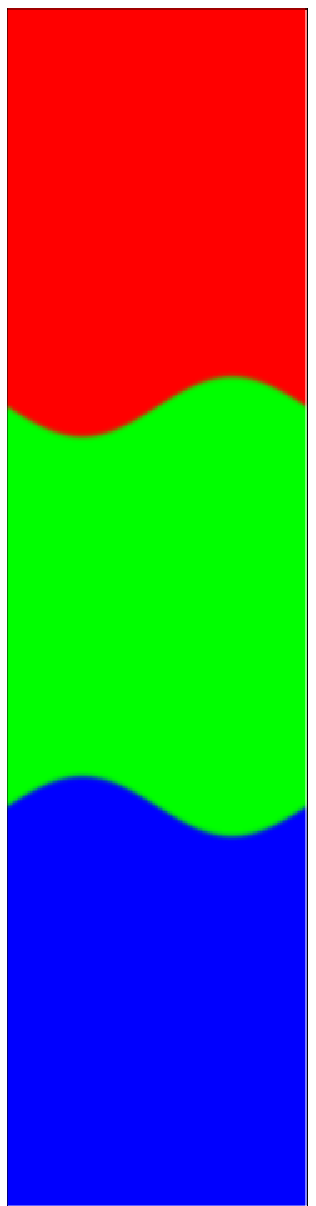}
	\end{minipage}}
	\subfigure[$T=2$]{
		\begin{minipage}{0.12\linewidth}
			\centering
			\includegraphics[width=.75in]{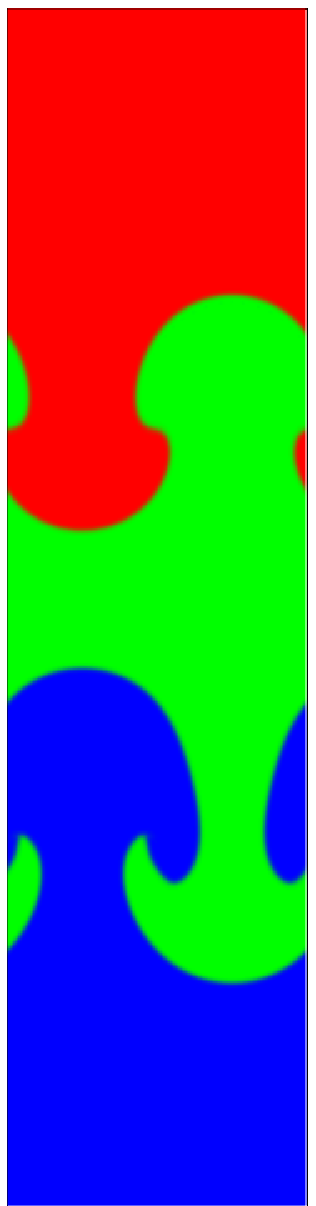}
	\end{minipage}}
	\subfigure[$T=4$]{
		\begin{minipage}{0.12\linewidth}
			\centering
			\includegraphics[width=.75in]{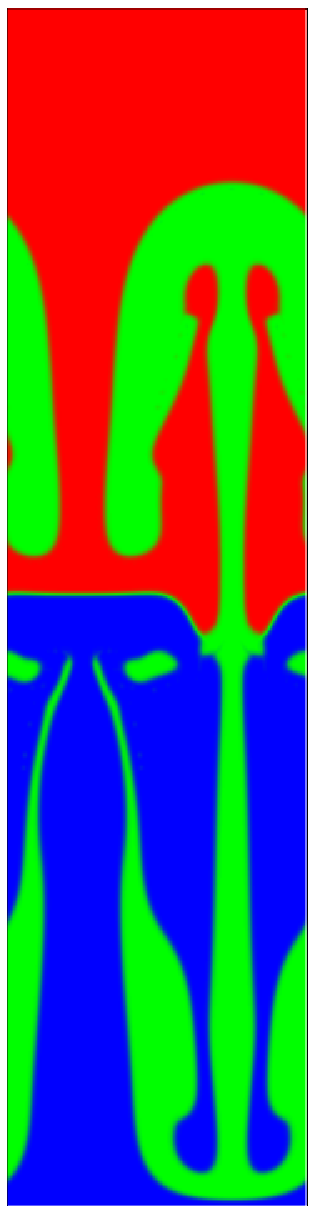}
	\end{minipage}}
	\subfigure[$T=5$]{
		\begin{minipage}{0.12\linewidth}
			\centering
			\includegraphics[width=.75in]{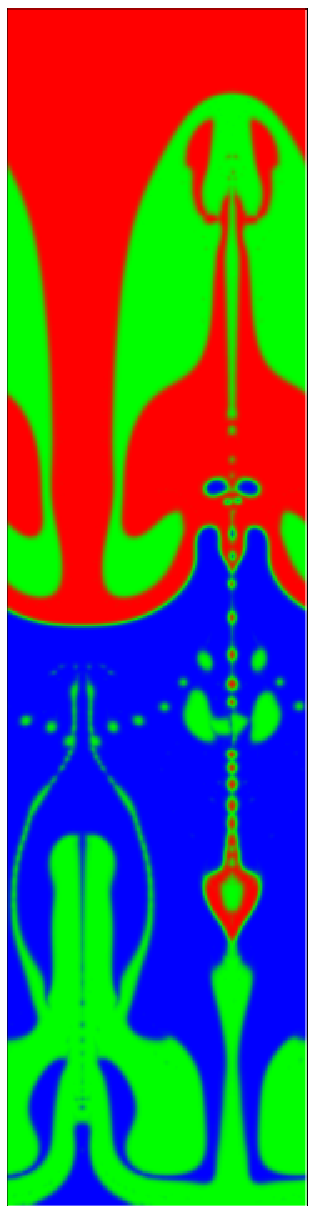}
	\end{minipage}}
	\subfigure[$T=10$]{
		\begin{minipage}{0.12\linewidth}
			\centering
			\includegraphics[width=.75in]{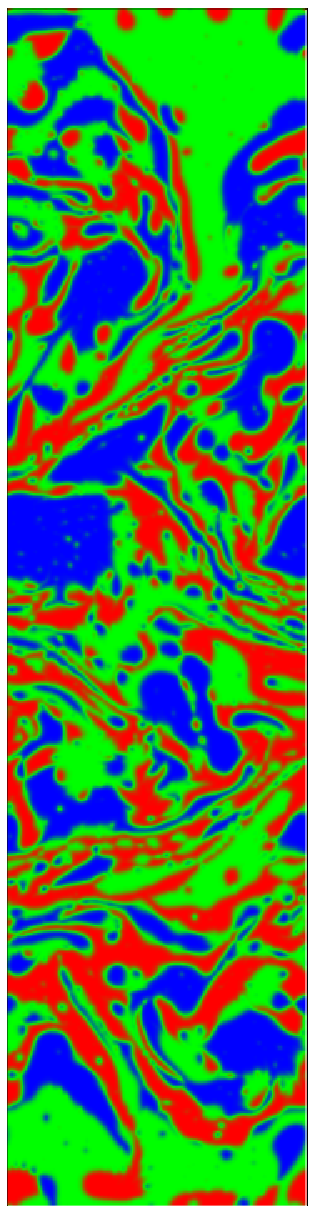}
	\end{minipage}}
	\subfigure[$T=25$]{
		\begin{minipage}{0.12\linewidth}
			\centering
			\includegraphics[width=.75in]{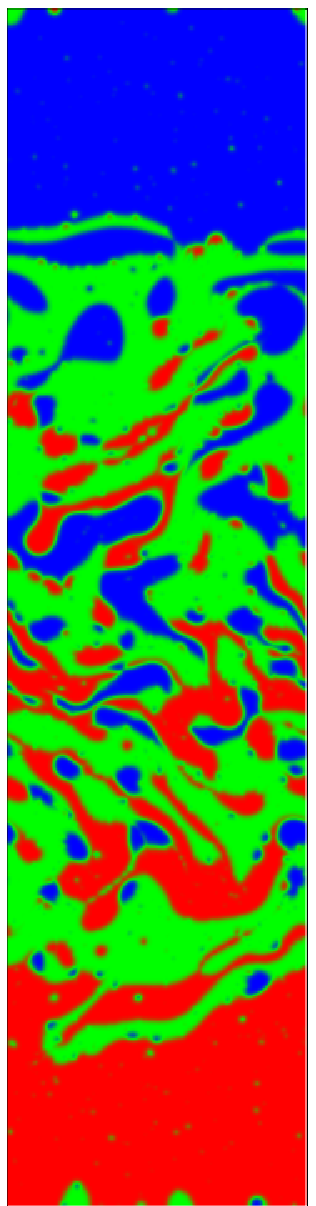}
	\end{minipage}}
	\subfigure[$T=100$]{
		\begin{minipage}{0.12\linewidth}
			\centering
			\includegraphics[width=.75in]{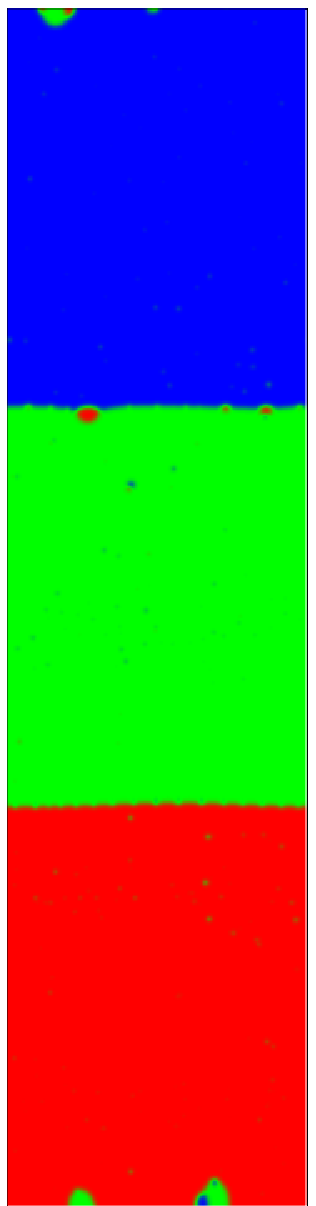}
	\end{minipage}}
	\caption{The snapshots of the RTI in ternary system.}
	\label{fig-RTI3}
\end{figure}

We now conduct some simulations for RTI in the ternary system. Compared to the binary case, the difference only lies in the initialization, and the phase-field variables here are given by
\begin{equation}
	\begin{aligned}
		\phi_1\left(x,y\right)&=0.5+0.5\tanh\frac{y-2d/3-0.1d\sin\left(2\pi x/d\right)}{\epsilon/2},\\
		\phi_2\left(x,y\right)&=\left[1-\phi_1\left(x,y\right)\right]\left[0.5+0.5\tanh\frac{y+2d/3+0.1d\sin\left(2\pi x/d\right)}{\epsilon/2}\right].
	\end{aligned}
\end{equation}
The time evolutions of the three fluids are plotted in Fig. \ref{fig-RTI3}, where two RTI mechanisms happen simultaneously with the different densities among three fluids. At the early stage, the initial perturbations grow, the lighter fluids rising while the heavier fluids falling. When the rising front of fluid 3 interacts with the falling spike tip of fluid 1, non-monotonic changes in the configuration occurs in this system. These three fluids eventually reach a essential equilibrium state due to the action of gravity. Actually, there are always some small droplets (or bubbles) in other fluids and they do not vanish in time, which illustrates the capacity of the current models in capturing small features.

\begin{figure}
	\centering
	\subfigure[]{
		\begin{minipage}{0.48\linewidth}
			\centering
			\includegraphics[width=3.0in]{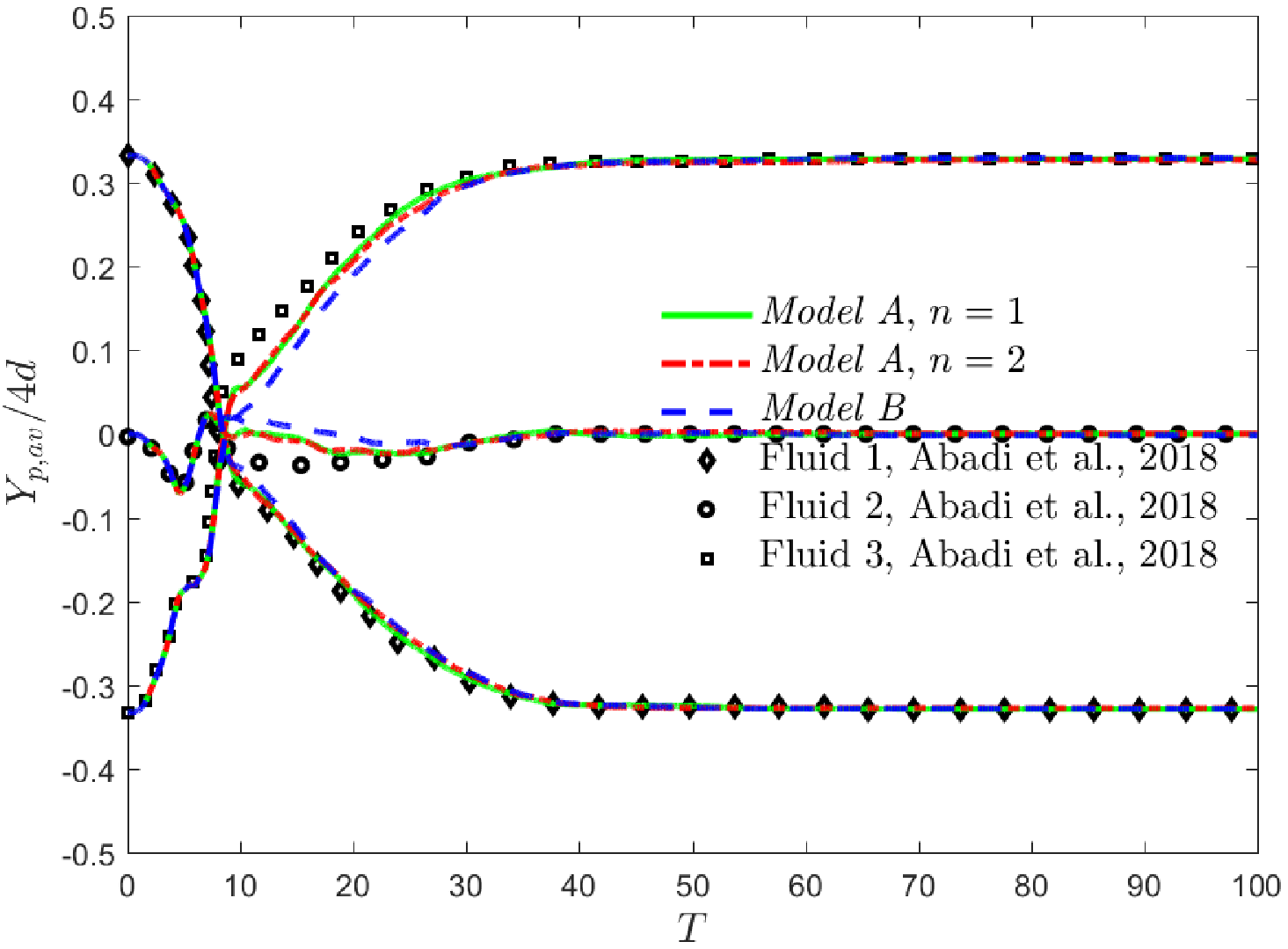}
	\end{minipage}}
	
	\subfigure[]{
		\begin{minipage}{0.48\linewidth}
			\centering
			\includegraphics[width=3.0in]{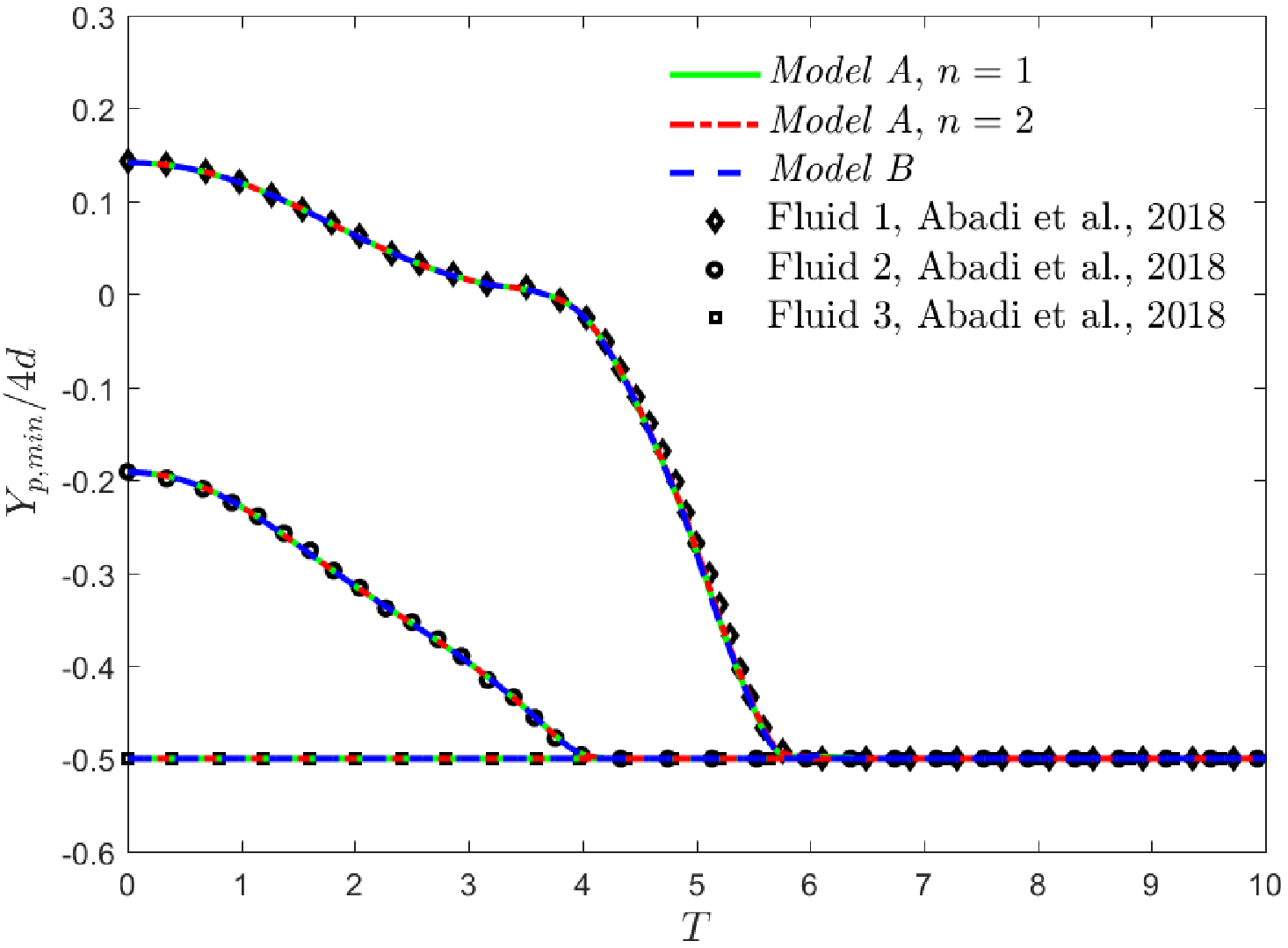}
	\end{minipage}}
	\subfigure[]{
		\begin{minipage}{0.48\linewidth}
			\centering
			\includegraphics[width=3.0in]{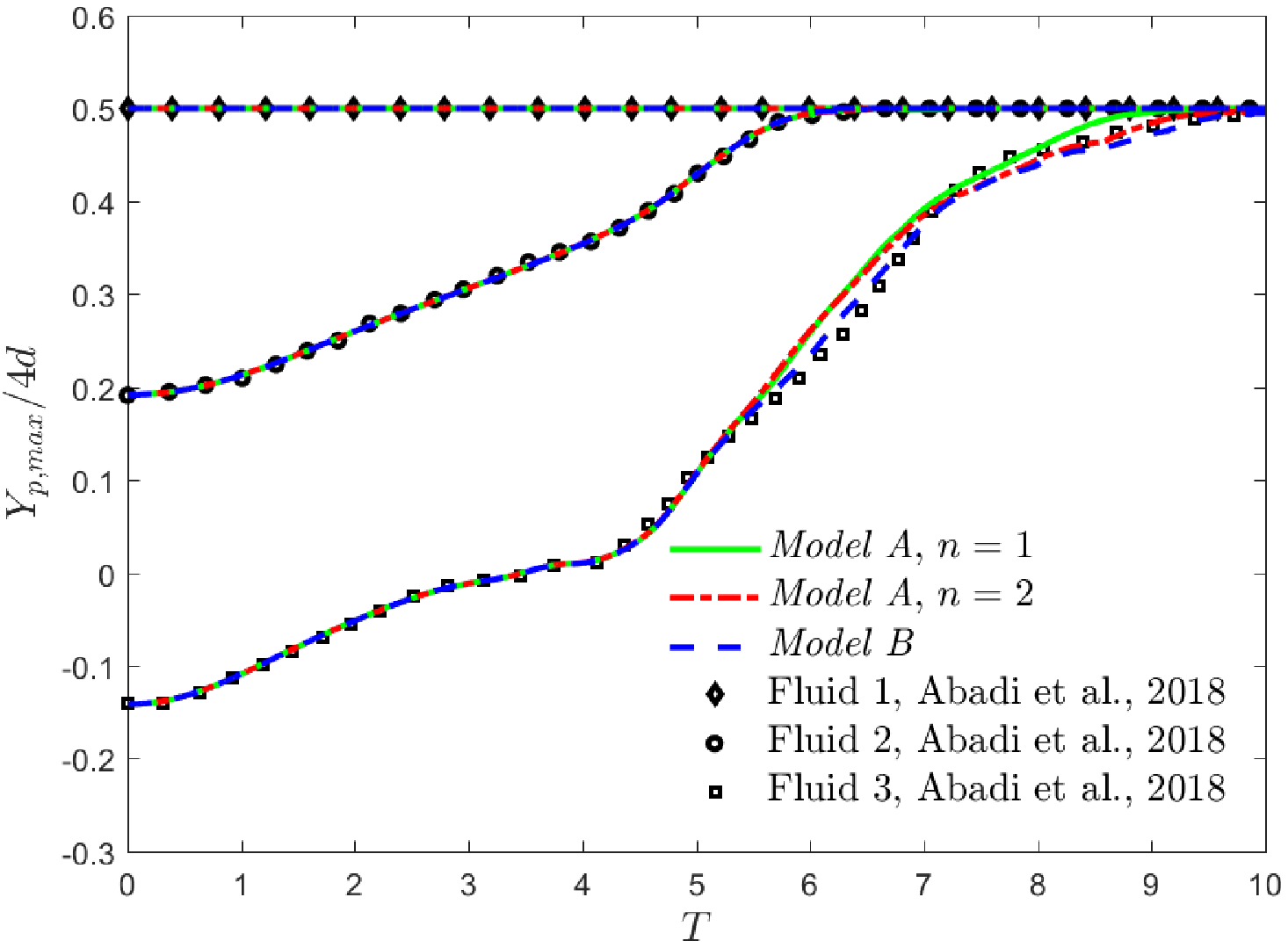}
	\end{minipage}}
	\caption{The time evolutions of the normalized locations of three fluids of the RTI in ternary system.}
	\label{fig-RTI3-com}
\end{figure} 

To quantify the RTI in a more complex ternary system, the average, the lowest interface, and the highest interface locations of each fluid component are considered,
\begin{equation}
	\begin{aligned}
		&Y_{p,av}=\frac{\sum_{\Omega}y\phi_p\left(\mathbf{x},t\right)}{\sum_{\Omega}\phi_p\left(\mathbf{x},t\right)},\quad p=1,2,3,\\
		&Y_{p,min}=\min\{y\in\Omega: \phi_p\left(\mathbf{x},t\right)\geq0.5\},\quad p=1,2,3,\\
		&Y_{p,max}=\max\{y\in\Omega: \phi_p\left(\mathbf{x},t\right)\geq0.5\},\quad p=1,2,3.
	\end{aligned}
\end{equation} 
The time evolutions of the above characteristic quantities are presented in Fig. \ref{fig-RTI3-com}, where the results of two AC models only have some differences in the early stage, and they are also consistent with those in Ref. \cite{Abadi2018JCP}. From this figure, one can observe that fluid 1 flows downwards while fluid 3 rises upwards due to the buoyancy force. Fluid 2 reaches the bottom and the top boundaries because of the interactions between different fluids. Although the spatial distributions of three fluids are erratic, their average locations do not change significantly with time at the late stage.

\subsection{Dam break}\label{dam}
We finally consider a dam break problem with complex topological changes, large density ratios and high Reynolds numbers. The material properties are set the same as those in Sec. \ref{lens} with the real gravity (listed in Table \ref{table-material}). The computational domain is $Lx\times Ly=0.4\times 0.1$, the water column is initially placed at the left bottom of the chamber, while the oil one is at the right with the same length $a=0.05$. For this problem, the initial phase-field variables are given by
\begin{equation}
	\begin{aligned}
		\phi_1\left(x,y\right)&=\begin{cases}
			0.5+0.5\tanh\frac{a-y}{\epsilon/2},\quad x\leq a-\epsilon, y\geq a-\epsilon,\\
			0.5+0.5\tanh\frac{a-x}{\epsilon/2},\quad x\geq a-\epsilon, y\leq a-\epsilon,\\
			0.5+0.5\tanh\frac{\epsilon-\sqrt{\left(x-a+\epsilon\right)^2+\left(y-a+\epsilon\right)^2}}{\epsilon/2},\quad x\geq a-\epsilon, y\geq a-\epsilon,\\
			1,\quad \text{otherwise},\\
		\end{cases}\\
		\phi_2\left(x,y\right)&=\begin{cases}
			0.5+0.5\tanh\frac{a-y}{\epsilon/2},\quad x\geq Lx-a+\epsilon, y\geq a-\epsilon,\\
			0.5+0.5\tanh\frac{x-Lx+a}{\epsilon/2},\quad x\leq Lx-a+\epsilon, y\leq a-\epsilon,\\
			0.5+0.5\tanh\frac{\epsilon-\sqrt{\left(x-Lx+a-\epsilon\right)^2+\left(y-a+\epsilon\right)^2}}{\epsilon/2},\quad x\leq Lx-a+\epsilon, y\geq a-\epsilon,\\
			1,\quad \text{otherwise}.
		\end{cases}\\
	\end{aligned}
\end{equation}

\begin{figure}
	\centering
	\subfigure[$T=0$]{
		\begin{minipage}{0.22\linewidth}
			\centering
			\includegraphics[width=1.45in]{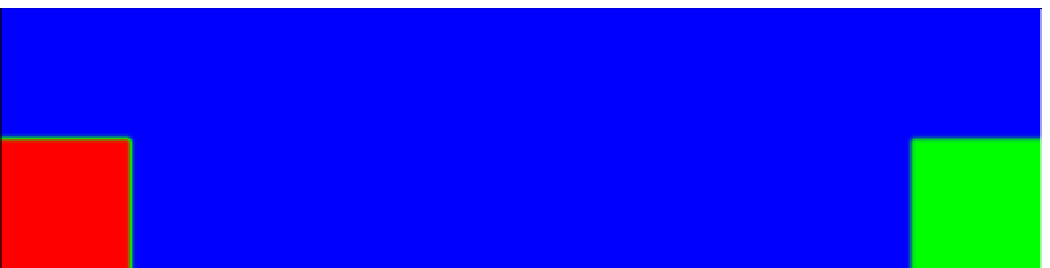}
	\end{minipage}}	
	\subfigure[$T=0.25$]{
		\begin{minipage}{0.22\linewidth}
			\centering
			\includegraphics[width=1.45in]{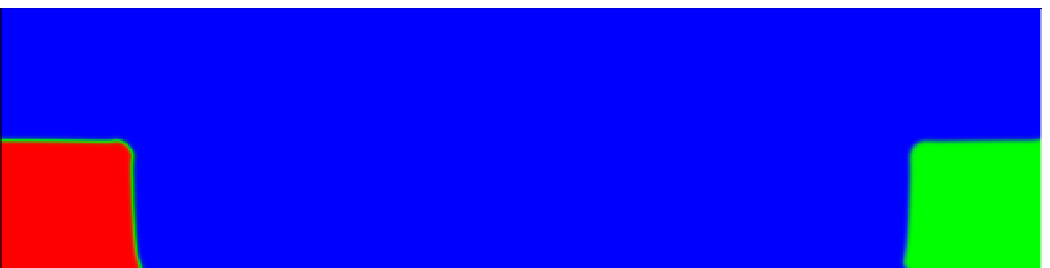}
	\end{minipage}}	
	\subfigure[$T=0.5$]{
		\begin{minipage}{0.22\linewidth}
			\centering
			\includegraphics[width=1.45in]{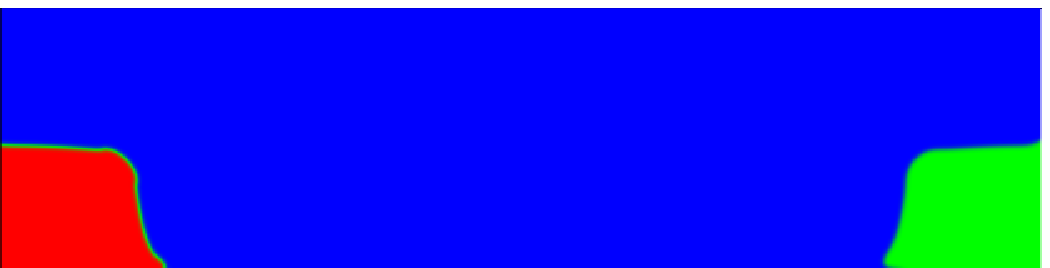}
	\end{minipage}}	
	\subfigure[$T=1$]{
		\begin{minipage}{0.22\linewidth}
			\centering
			\includegraphics[width=1.45in]{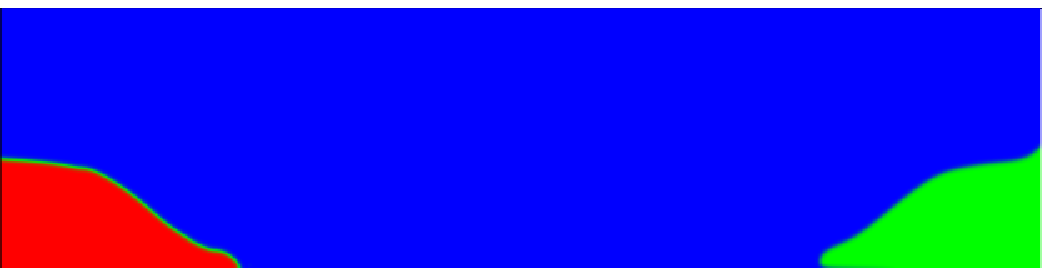}
	\end{minipage}}	
	
	\subfigure[$T=1.25$]{
		\begin{minipage}{0.22\linewidth}
			\centering
			\includegraphics[width=1.45in]{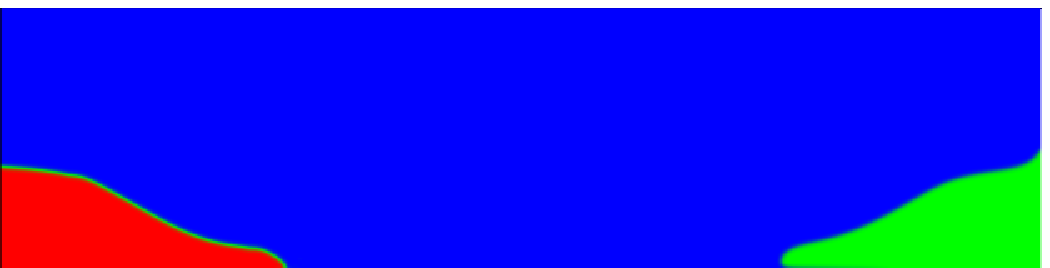}
	\end{minipage}}	
	\subfigure[$T=1.5$]{
		\begin{minipage}{0.22\linewidth}
			\centering
			\includegraphics[width=1.45in]{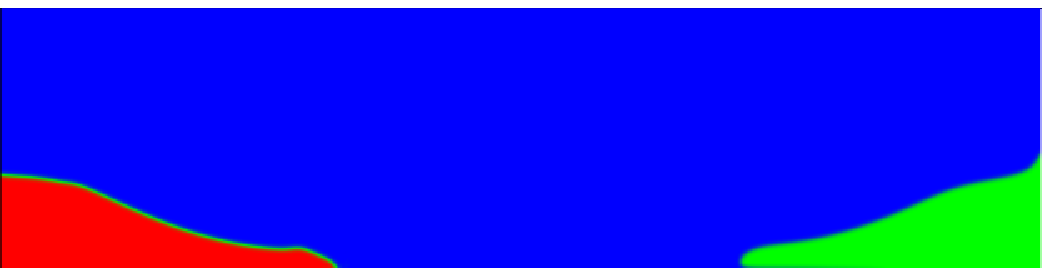}
	\end{minipage}}	
	\subfigure[$T=2$]{
		\begin{minipage}{0.22\linewidth}
			\centering
			\includegraphics[width=1.45in]{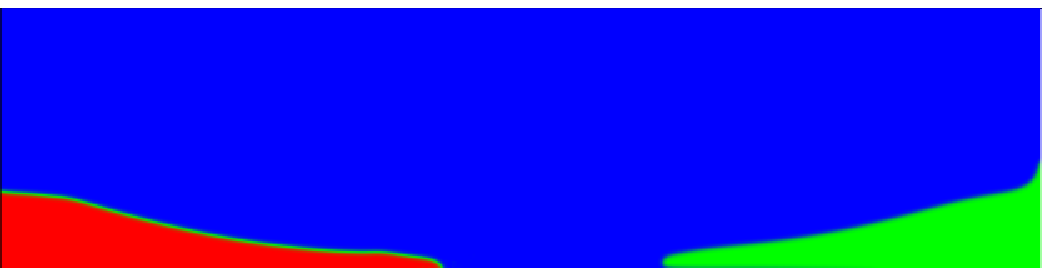}
	\end{minipage}}	
	\subfigure[$T=2.25$]{
		\begin{minipage}{0.22\linewidth}
			\centering
			\includegraphics[width=1.45in]{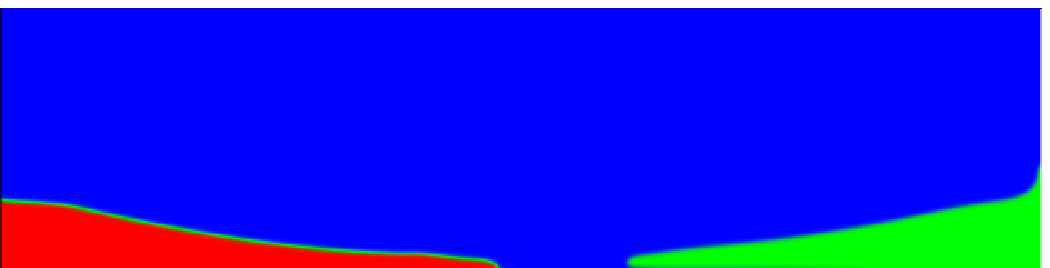}
	\end{minipage}}	
	\caption{The snapshots of the ternary dam break at the early stage.}
	\label{fig-dam3-1}
\end{figure}
\begin{figure}
	\centering
	\subfigure[]{
		\begin{minipage}{0.48\linewidth}
			\centering
			\includegraphics[width=3.0in]{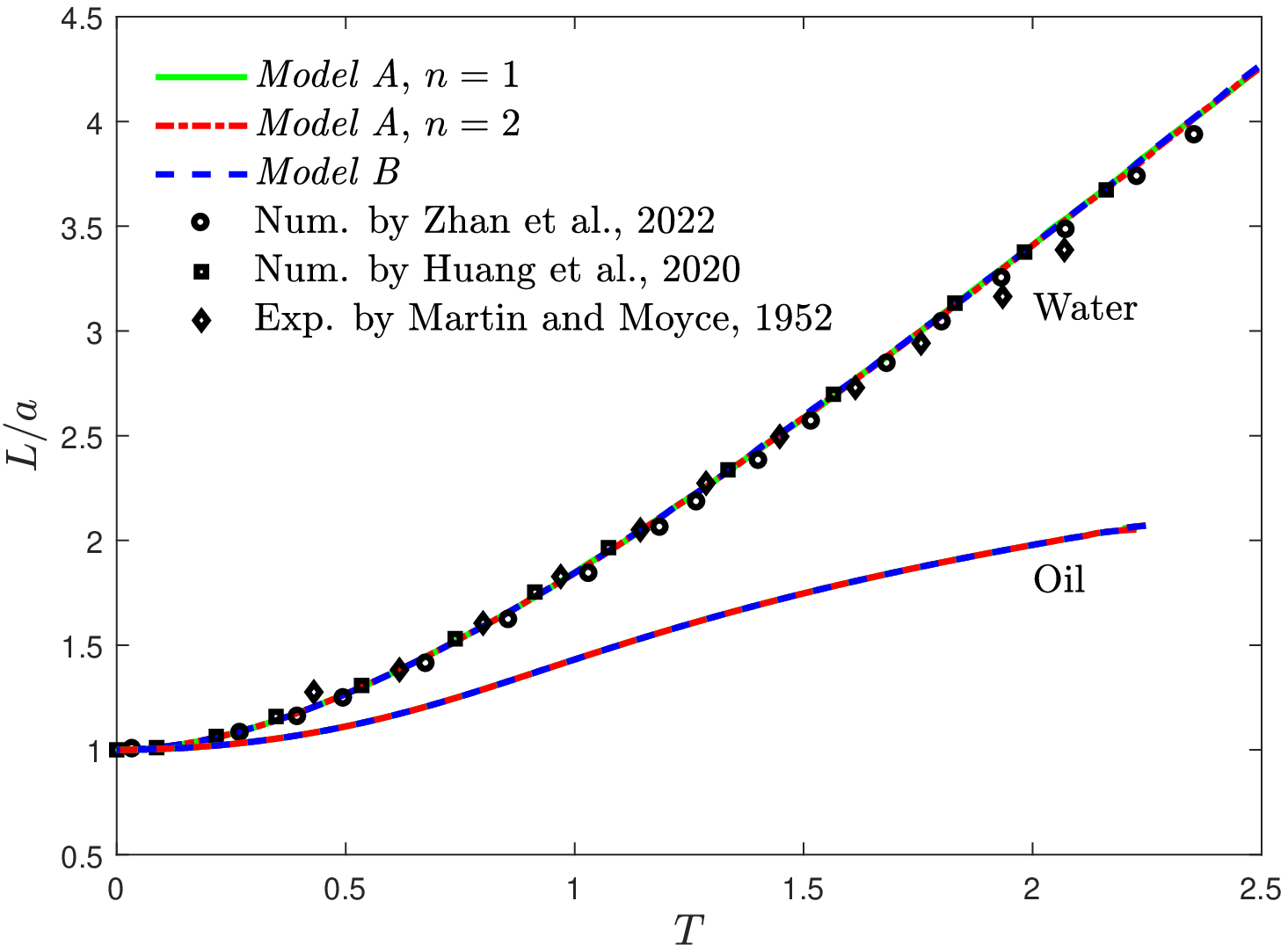}
	\end{minipage}}	
	\subfigure[]{
		\begin{minipage}{0.48\linewidth}
			\centering
			\includegraphics[width=3.0in]{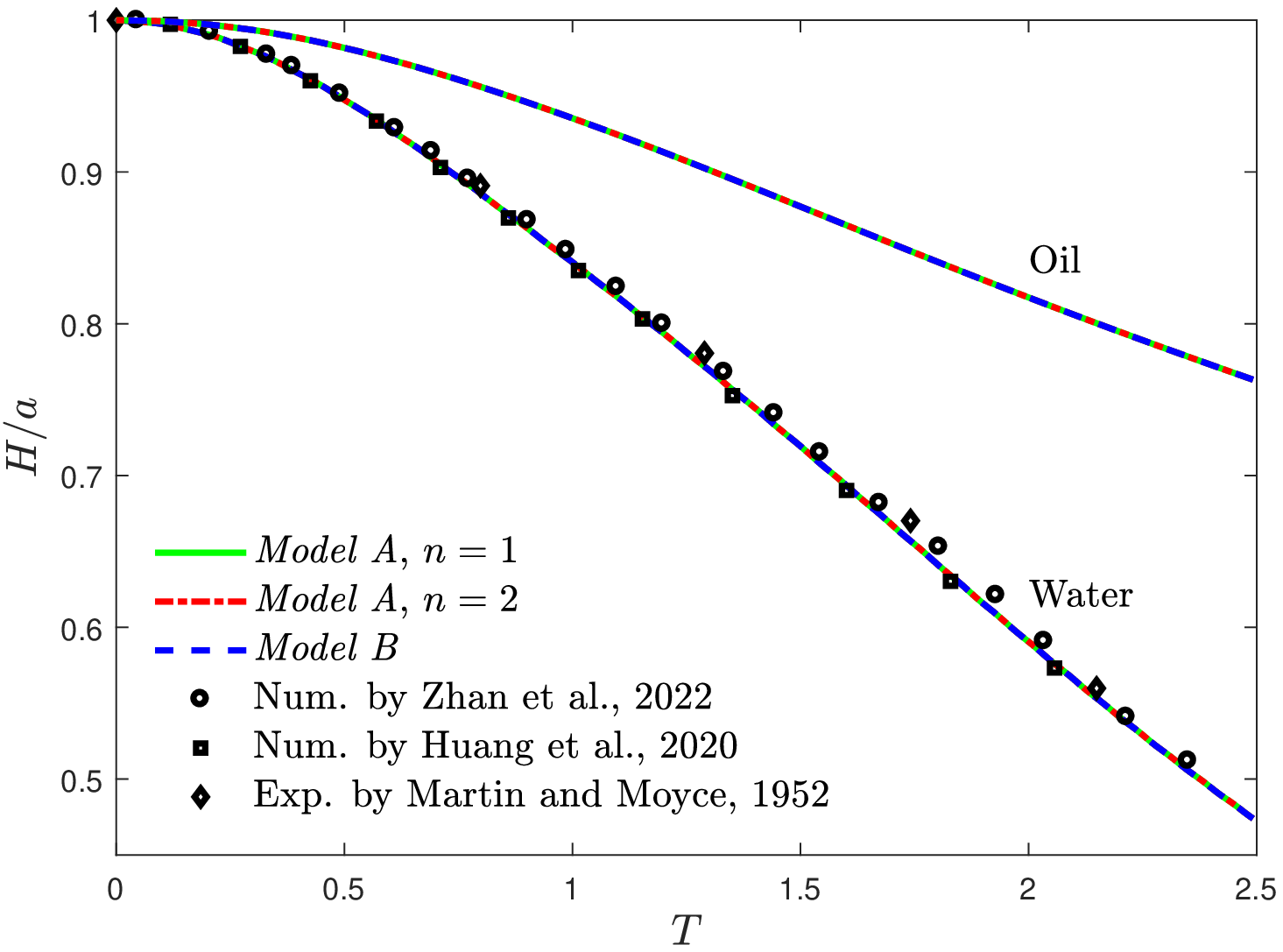}
	\end{minipage}}	
	\caption{Front and height of the water and oil columns in the ternary dam break problem [(a) The normalized location of the front $L/a$, (b) the normalized location of the height $H/a$].}
	\label{fig-dam3-HL}
\end{figure}
\begin{figure}
	\centering
	\subfigure[\emph{Model A}, $n=1$]{
		\begin{minipage}{0.48\linewidth}
			\centering
			\includegraphics[width=3.0in]{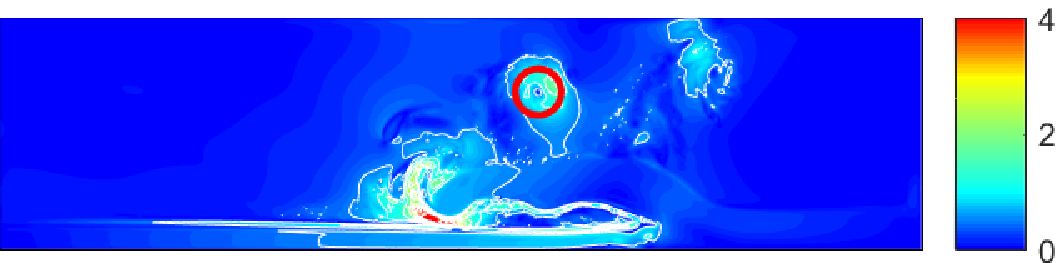}
	\end{minipage}}	
	\subfigure[\emph{Model A}, $n=2$]{
		\begin{minipage}{0.48\linewidth}
			\centering
			\includegraphics[width=3.0in]{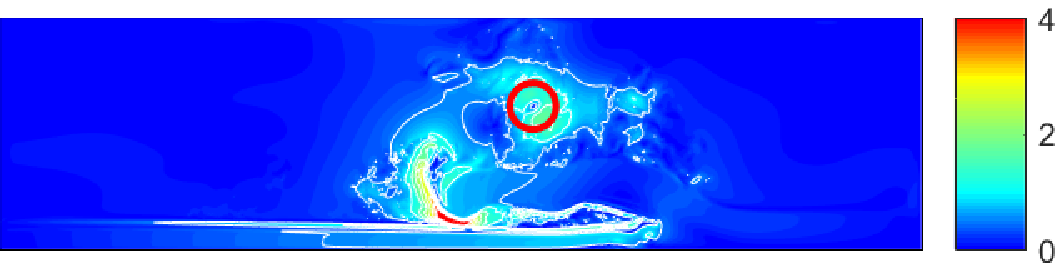}
	\end{minipage}}	
	\caption{The snapshots of velocity intensity $|\mathbf{u}|$ superimposed with its contour lines at the time before \emph{Model A} becoming unstable.}
	\label{fig-dam3-com}
\end{figure}
\begin{figure}
	\centering
	\subfigure[$T=2.5$]{
		\begin{minipage}{0.22\linewidth}
			\centering
			\includegraphics[width=1.45in]{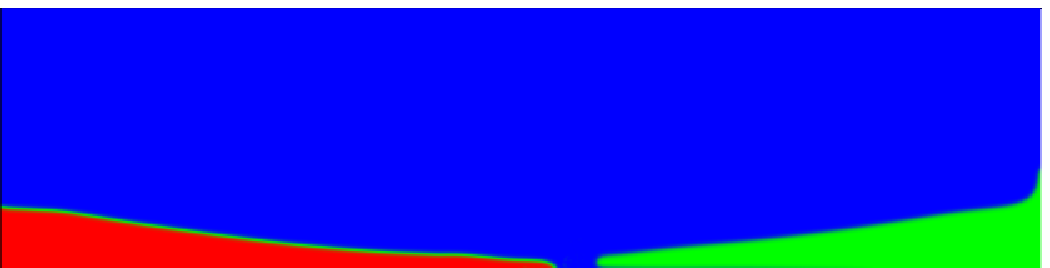}
	\end{minipage}}	
	\subfigure[$T=3$]{
		\begin{minipage}{0.22\linewidth}
			\centering
			\includegraphics[width=1.45in]{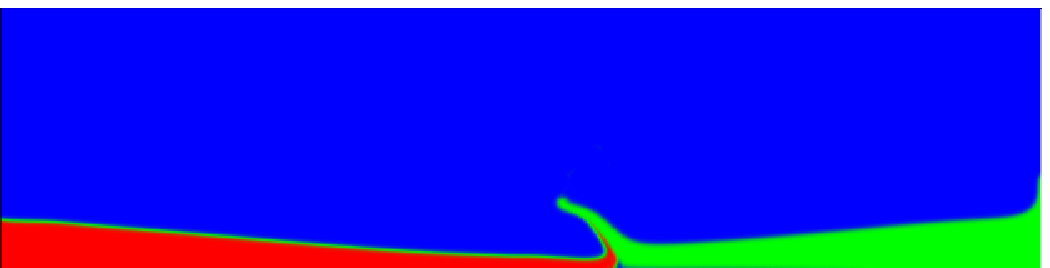}
	\end{minipage}}	
	\subfigure[$T=3.5$]{
		\begin{minipage}{0.22\linewidth}
			\centering
			\includegraphics[width=1.45in]{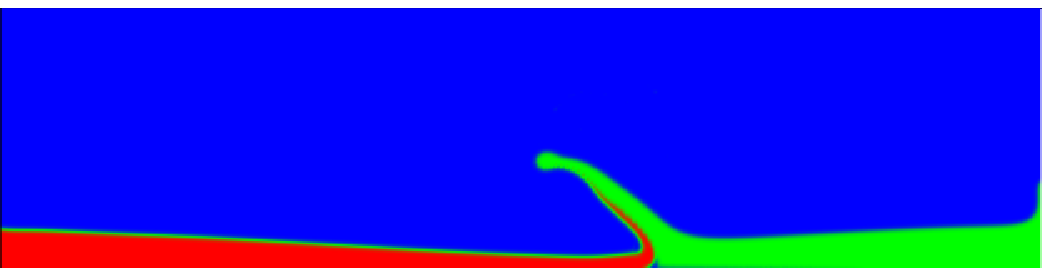}
	\end{minipage}}	
	\subfigure[$T=4$]{
		\begin{minipage}{0.22\linewidth}
			\centering
			\includegraphics[width=1.45in]{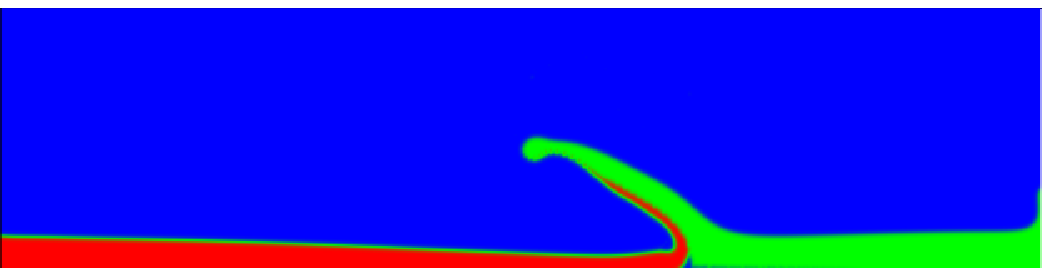}
	\end{minipage}}	
	
	\subfigure[$T=4.5$]{
		\begin{minipage}{0.22\linewidth}
			\centering
			\includegraphics[width=1.45in]{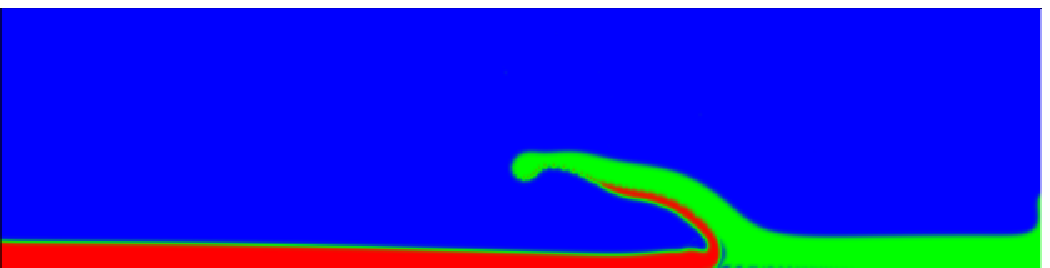}
	\end{minipage}}	
	\subfigure[$T=5$]{
		\begin{minipage}{0.22\linewidth}
			\centering
			\includegraphics[width=1.45in]{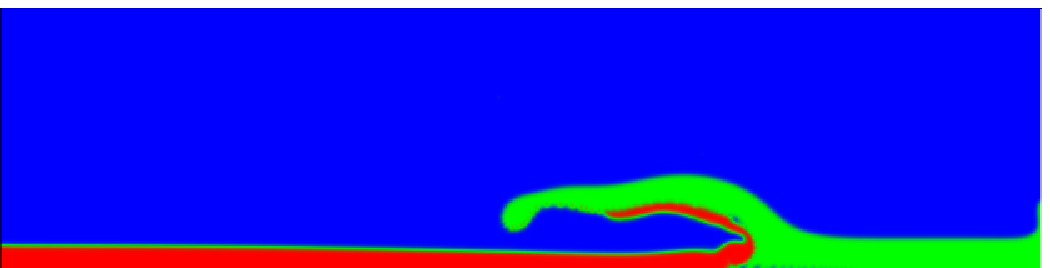}
	\end{minipage}}	
	\subfigure[$T=5.5$]{
		\begin{minipage}{0.22\linewidth}
			\centering
			\includegraphics[width=1.45in]{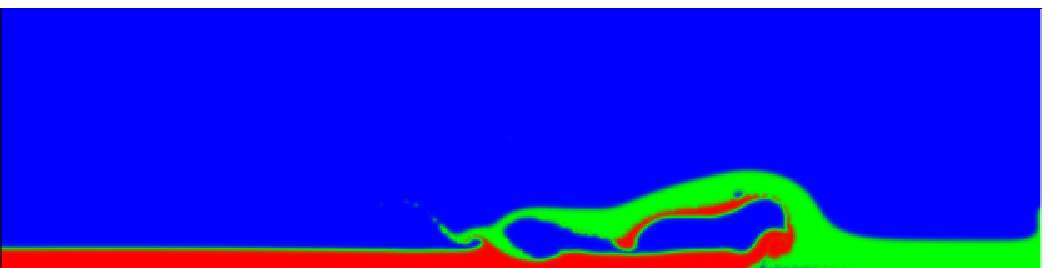}
	\end{minipage}}	
	\subfigure[$T=6$]{
		\begin{minipage}{0.22\linewidth}
			\centering
			\includegraphics[width=1.45in]{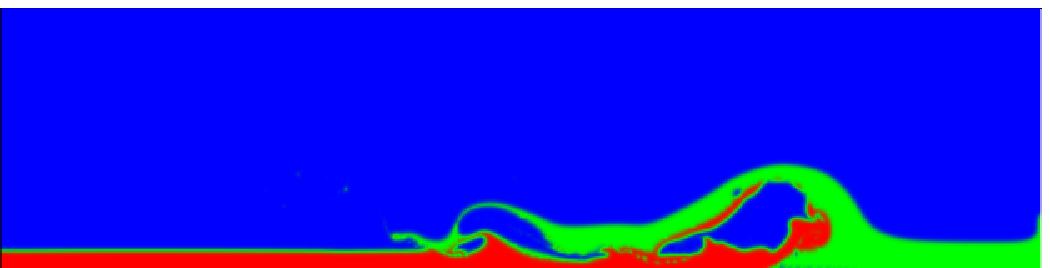}
	\end{minipage}}	
	
	\subfigure[$T=6.5$]{
		\begin{minipage}{0.22\linewidth}
			\centering
			\includegraphics[width=1.45in]{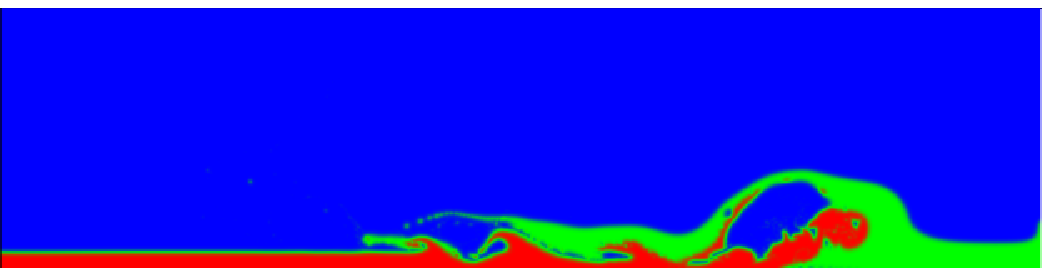}
	\end{minipage}}	
	\subfigure[$T=7$]{
		\begin{minipage}{0.22\linewidth}
			\centering
			\includegraphics[width=1.45in]{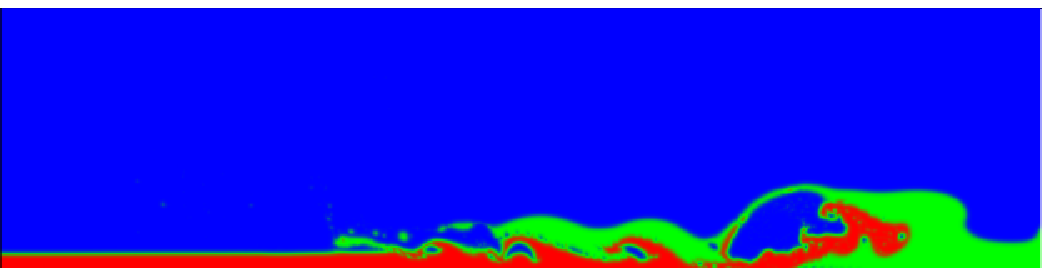}
	\end{minipage}}	
	\subfigure[$T=7.5$]{
		\begin{minipage}{0.22\linewidth}
			\centering
			\includegraphics[width=1.45in]{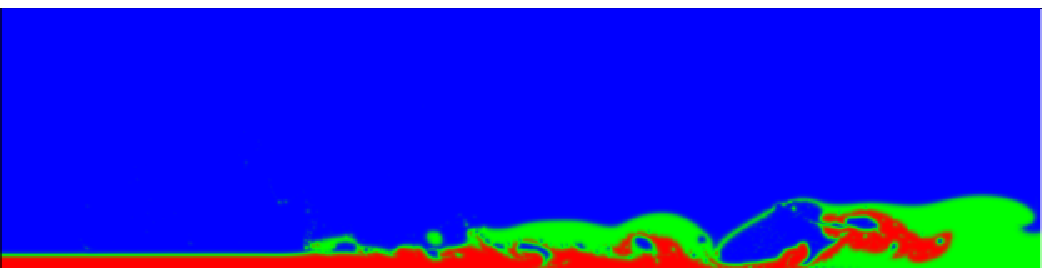}
	\end{minipage}}	
	\subfigure[$T=8$]{
		\begin{minipage}{0.22\linewidth}
			\centering
			\includegraphics[width=1.45in]{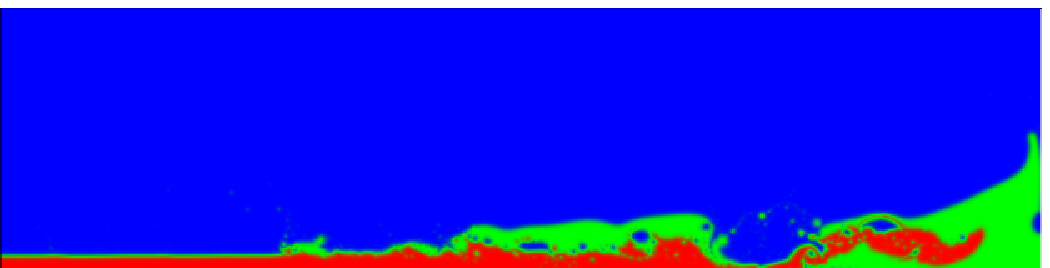}
	\end{minipage}}	
	
	\subfigure[$T=8.5$]{
		\begin{minipage}{0.22\linewidth}
			\centering
			\includegraphics[width=1.45in]{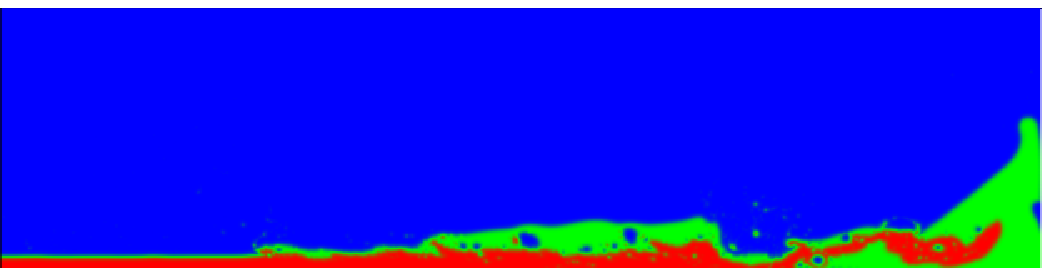}
	\end{minipage}}	
	\subfigure[$T=9$]{
		\begin{minipage}{0.22\linewidth}
			\centering
			\includegraphics[width=1.45in]{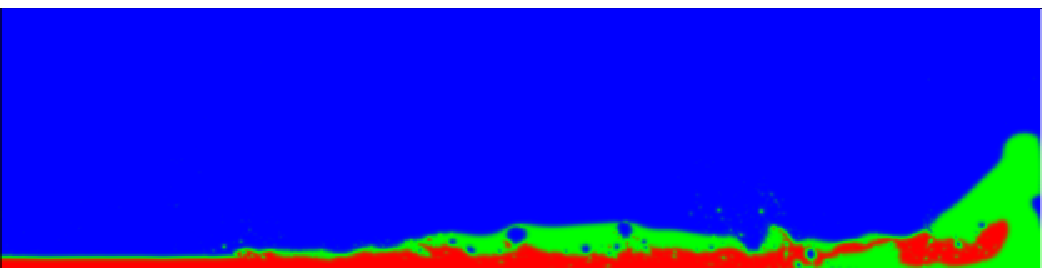}
	\end{minipage}}	
	\subfigure[$T=9.5$]{
		\begin{minipage}{0.22\linewidth}
			\centering
			\includegraphics[width=1.45in]{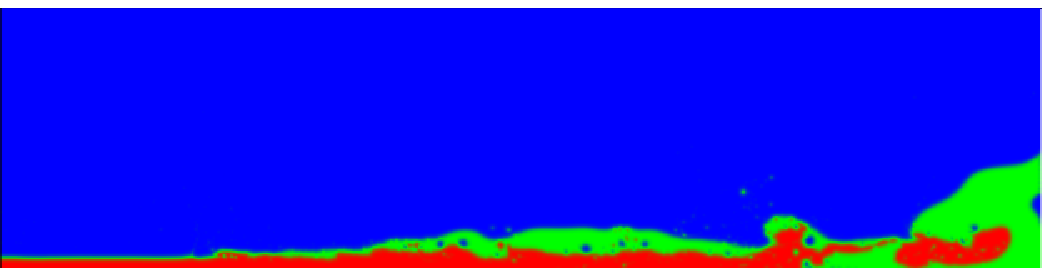}
	\end{minipage}}	
	\subfigure[$T=10$]{
		\begin{minipage}{0.22\linewidth}
			\centering
			\includegraphics[width=1.45in]{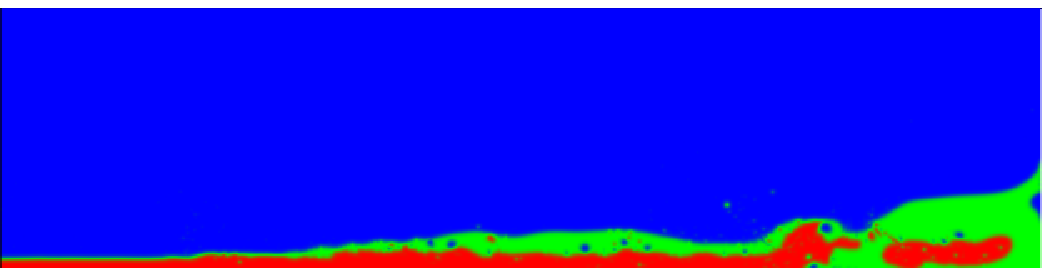}
	\end{minipage}}	
	\caption{The snapshots of the ternary dam break in the long-time evolution (\emph{Model B}).}
	\label{fig-dam3-2}
\end{figure}

In our simulations, the lattice spacing and time step are set as $\Delta x=1.25\times10^{-4}$ and $\Delta t=\Delta x/25$. Under the influence of gravity, both the water and oil columns will collapse. At the early stage of the evolution (see Fig. \ref{fig-dam3-1} with the dimensionless time $T=t/\sqrt{a/g}<2.5$), the water and oil spread separately on the bottom wall without interaction due to the large value of horizontal distance $Lx$. In this case, the locations of the front $L$ and the height $H$ can be compared with the results of two-phase (water-air) dam break in Refs. \cite{Martin1952MPS,Huang2020JCPa,Zhan2022PRE}. As shown in Fig. \ref{fig-dam3-HL}, the present results of two AC models agree well with the two-phase experimental measurements and numerical data, and additionally, one can also find that the oil front and height move slower than those of water due to the larger viscosity of the oil.

When the fronts of water and oil meet with each other, the complex dynamics of three phases happens in the chamber. In this stage, \emph{Model A} becomes unstable at about $T=5$, which may caused by the singularity of the velocity field in the area of background phase, as labeled in Fig. \ref{fig-dam3-com}. We also plot the evolution of the three-phase dam break simulated by \emph{Model B} in Fig. \ref{fig-dam3-2}, where the time is varied from $T=2.5$ to $T=10$. We note that similar to the processes shown in Ref. \cite{Huang2022JCAM}, the oil front is squeezed upward by the water, and the water continue to move to the right wall. In the meanwhile, the squeezed oil front collapses again onto the water, breaks up into small droplets and filaments, and very strong interaction between water, oil, and air occurs. From above discussion, one can conclude that the \emph{Model B} is more stable in the study of realistic complex multiphase flows.

\section{Conclusions}\label{Conclusion}
In this paper, we conduct a detailed comparison of two conservative AC models for $N$-phase flows, which are developed from different point of view but have some relations. These two AC models are first compared theoretically, and the results show that they are equivalent to each other under the conditions of $n=1$ and some proper approximations of the gradient norms. To further numerically compare these two models and investigate the effects of the approximations, we propose a consistent and conservative LB method for the phase-field model and conduct some simulations. Three benchmark problems with analytical solutions, i.e., the static droplets, the spreading of a compound droplet on a solid wall, and the spreading of a liquid lens, are used to test the accuracy of present LB method. Then, the ternary-fluid RTI and the ternary dam break problems with long-time dynamics are adopted. The numerical results illustrate that for the simple problems with appropriate parameters, the \emph{Model A} and \emph{Model B} are both numerically stable, and can give accurate results. However, the \emph{Model B} has a better capability for the practical water-oil-air problems with the long-time dynamics, whereas the \emph{Model A} may be inaccurate and unstable. This conclusion also indicates that the approximations of the gradient norms may have some negative effects in present LB simulations with $n=1$. For the value of $n$ in \emph{Model A}, we only consider $n=1$ and 2 in this work, and the performance of $n=1$ seems better than that of $n=2$, while the optimal value of $n$ needs a further study in a future work.

\section*{Acknowledgments}
This research has been supported by the National Natural Science Foundation of China under Grants No. 12072127 and No 51836003. The computation was completed on the HPC Platform of Huazhong University of Science and Technology. We also thank for Prof. Lin Zheng for the useful discussion. 

\bibliographystyle{elsarticle-num} 
\bibliography{reference}
\end{document}